\documentclass[aps,pra,superscriptaddress,reprint,floatfix,showpacs]{revtex4-1}

\usepackage{bm}
\usepackage{graphicx}
\usepackage{amsmath}

\usepackage{amsmath,amsfonts,amssymb}

\begin{document}

\title{Self-consistent tight-binding description of Dirac points moving and merging in two dimensional optical lattices}

\author{Julen Iba\~nez-Azpiroz}
\affiliation{Depto. de F\'isica de la Materia Condensada, Universidad del Pais Vasco, UPV/EHU, 48080 Bilbao, Spain}
\affiliation{Donostia International Physics Center (DIPC), 20018 Donostia, Spain}

\author{Asier Eiguren}
\affiliation{Depto. de F\'isica de la Materia Condensada, Universidad del Pais Vasco, UPV/EHU, 48080 Bilbao, Spain}
\affiliation{Donostia International Physics Center (DIPC), 20018 Donostia, Spain}

\author{Aitor Bergara}
\affiliation{Depto. de F\'isica de la Materia Condensada, Universidad del Pais Vasco, UPV/EHU, 48080 Bilbao, Spain}
\affiliation{Donostia International Physics Center (DIPC), 20018 Donostia, Spain}
\affiliation{Centro de F\'{i}sica de Materiales CFM, Centro Mixto CSIC-UPV/EHU, 20018 Donostia, Spain}

\author{Giulio Pettini}
\affiliation{Dipartimento di Fisica e Astronomia, Universit\`a di Firenze,
and INFN, 50019 Sesto Fiorentino, Italy}

\author{Michele Modugno}
\affiliation{\mbox{Depto. de F\'isica Te\'orica e Hist. de la Ciencia, Universidad del Pais Vasco UPV/EHU, 48080 Bilbao, Spain}}
\affiliation{IKERBASQUE, Basque Foundation for Science, 48011 Bilbao, Spain}

\begin{abstract}
We present an accurate \textit{ab initio} tight-binding model, capable of describing the dynamics 
of Dirac points in tunable honeycomb optical lattices following a recent experimental realization 
[L. Tarruell \textit{et al.}, Nature \textbf{483}, 302 (2012)].
Our scheme is based on first-principle maximally localized Wannier functions for composite bands.
The tunneling coefficients are calculated for different lattice configurations, and the spectrum properties are well reproduced with high accuracy.
In particular, we show which tight binding description is needed in order
to accurately reproduce the position of Dirac points and the dispersion law close to their merging, for different laser intensities.
\end{abstract}

\date{\today}

\maketitle

\section{Introduction.}

The possibility to simulate graphene-like structures and investigate the
physics of Dirac points with ultra cold atoms in optical lattices is attracting an increasing interest in the literature \cite{zhu2007,wu2007,wunsch2008,lee2009,montambaux2009,soltan2011,sun2012,tarruell2012,lim2012,uehlinger2012}.
Recently, Tarruel \textit{et al.} \cite{tarruell2012} have reported the creation and manipulation of Dirac points in a tunable honeycomb optical lattice, exploring the topological transition occurring at the merging of Dirac points, and comparing their experimental results with \textit{ab initio} calculations of the Bloch spectrum.
The same experiment has also been interpreted by means of an universal tight-binding Hamiltonian
defined on a \textit{square} lattice \cite{lim2012}, that describes the merging of Dirac points and the corresponding topological transition between a semi-metallic phase and a band insulator \cite{montambaux2009}.
Though this model remarkably captures all the relevant physics, its connection with the optical lattice parameters is  indirect and has some limitations, as it relies on a fit of the parameters of the universal Hamiltonian to the two lowest lying energy bands in the vicinity of the Dirac cones \cite{lim2012,uehlinger2012}.

In this paper, we present a comprehensive scheme based on composite maximally localized Wannier functions (MLWFs) \cite{marzari1997} for constructing an \textit{ab initio}  tight-binding model corresponding to the \textit{tunable honeycomb} potential with two minima per unit cell as described in Ref. \cite{tarruell2012}.
The MLWFs are obtained by means of a gauge transformation that minimizes their real space spread, and are
routinely employed in condensed matter physics \cite{marzari2012}. As recently demonstrated, these functions represent also 
an optimal tool for
constructing tight-binding models for ultra cold atoms in optical lattices \cite{modugno2012,ibanez2013,cotugno2013}, 
as they allow for an optimized mapping
of the system hamiltonian onto the discrete model defined on a lattice.
They allow for an ab initio calculation of the tight-binding parameters, with a fine control over next to leading corrections.

The paper is organized as follows. In section \ref{sec:mlwf}, we review the general approach for mapping a continuous many-body Hamiltonian onto a discrete tight-binding model by means of MLWFs. In section \ref{sec:honeycomb} we apply this scheme in the tunable two-dimensional honeycomb lattice of Ref. \cite{tarruell2012}, discussing the general structure of the associated Bravais lattice in direct and reciprocal space, and presenting the tight-binding expansion up to the third-nearest neighbors. 
Here we also give explicit numerical results for the calculated MLWFs, the inferred tunneling coefficients and the Bloch band structure.
Then, in section IV, we make use of the tight-binding Hamiltonian in reciprocal space - expressed in terms of tunneling coefficients and functions depending on the geometry of the associated lattice - for discussing the behavior of the Dirac points as a function of the lattice parameters. There, we also discuss the dispersion relation close to the merging of Dirac
points, refining and improving the analysis of Refs. \cite{montambaux2009,lim2012}.
In addition, we discuss the effect of parity breaking, that destroys the degeneracy of the two potential minima, providing a finite Dirac mass. 
Some accessory but important details are included in the Appendices, covering  
the tight-binding expansion, the gauge dependence of the results, and the numerical application. 

\section{Tight binding expansion and MLWF\lowercase{s}}
\label{sec:mlwf}

Let us consider a many-body system of bosonic or fermionic particles,
described by the field operator $\hat{\psi}(\bm{r})$. In particular, as the physics of Dirac points is
determined by the single-particle spectrum, here we will consider the 
non-interacting many-body Hamiltonian
\begin{equation}
\hat{\cal{H}}_{0}=\int d\bm{r}~{\hat{\psi}}^\dagger(\bm{r})\hat{H}_{0}{\hat{\psi}}(\bm{r})
\label{eq:genham}
\end{equation}
with ${\hat{H}}_{0}=-(\hbar^{2}/2m)\nabla^{2} + V(\bm{r})$ and
the lattice periodic potential
$V(\bm{r})= V(\bm{r}+\bm{R})$, where $\bm{R}$ belongs to the associated Bravais lattice.
We remark that the optimal Wannier basis is solely determined by the single particle spectrum, and even though the inclusion of an interaction would be straightforward \cite{jaksch1998,modugno2012,cotugno2013}, the derivations following the next lines would not be affected.

The Hamiltonian (\ref{eq:genham}) can be
conveniently mapped onto a discrete lattice corresponding to
the minima of the potential $V(\bm{r})$ 
by expanding the field operator in terms of a set of functions $\{w_{\bm{j}\nu}(\bm{r})\}$
localized at each minimum,
\begin{equation}
\hat{\psi}(\bm{r})\equiv \sum_{\bm{j}\nu}{\hat{a}}_{\bm{j}\nu}w_{\bm{j}\nu}(\bm{r})
\label{eq:psiexpf}
\end{equation}
where $\nu$ is a band index, and 
$\hat{a}_{\bm{j}\nu}^{\dagger}$ ($\hat{a}_{\bm{j}\nu}$)
represent the creation (destruction) operators of a single particle in the $\bm{j}$-th cell,
satisfying the usual commutation (or anti commutation) rules 
following from those for the field $\hat{\psi}$.

The MLWFs, introduced by Marzari and Vanderbilt in \cite{marzari1997}, 
are obtained trough a unitary transformation of the Bloch eigenstates
\begin{eqnarray}
\langle x|w_{\bm{j}\nu}\rangle&=&\frac{1}{\sqrt{V_{\cal B}}}
\int_{\cal B} \!\!d\bm{k} ~e^{-i\bm{k}\bm{R}_{\bm{j}}}\sum_{m=1}^{N}U_{\nu m}(\bm{k})\langle x|\psi_{m\bm{k}}\rangle,
\label{eq:mlwfs}
\end{eqnarray}
with $V_{\cal B}$ the volume of the first Brillouin zone and $U\in U(N)$ a gauge transformation
which obeys periodicity conditions in order to preserve the Bloch theorem. 
This gauge transformation is obtained through the minimization of the 
Marzari-Vanderbilt localization functional
$\Omega=\sum_{\nu}\left[\langle \bm{r}^2\rangle_{\nu}-\langle \bm{r}\rangle_{\nu}^{2}\right]$ \cite{marzari1997}.
The resulting MLWFs posses the desirable property
of being exponentially localized in real space, \cite{brouder2007,panati2011}  
thus constituting an ideal basis for tight-binding models \cite{ibanez2013}.  
In this article we consider the MLWFs for composite bands (N$>$1)
since we are interested in geometries where the  Wigner-Seitz cell has a non trivial basis.
This allows each MLWF to be centered on a 
single potential minimum inside the elementary cell, 
in contrast to single-band Wannier functions \cite{modugno2012,ibanez2013,vaucher2007}.
For this work, the MLWFs have been computed by means of the WANNIER90 code \cite{mostofi2008,marzari2012}
and a modified version of the QUANTUM-ESPRESSO package \cite{espresso}
adapted to the case of an optical lattice \cite{ibanez2013}.
We also mention that other methods for specific cases have been recently proposed \cite{modugno2012,cotugno2013}.

The Hamiltonian (\ref{eq:genham}) can be written in terms of Wannier states $|w_{\bm{j}\nu}\rangle$ as
\begin{equation}
{\hat{\cal{H}}}_0 = \sum_{\nu\nu'=A,B}\sum_{\bm{j,j'}}{\hat{a}}^{\dagger}_{\bm{j}\nu}{\hat{a}}_{\bm{j}'\nu'}
\langle w_{\bm{j}\nu}|{\hat{H}}_0|w_{\bm{j'}\nu '}\rangle
\label{singparth0}
\end{equation}
where the matrix elements $\langle w_{\bm{j}\nu}|{\hat{H}}_0|w_{\bm{j'}\nu '}\rangle$ depend only on $\bm{i}=\bm{j'-j}$ due to the translational invariance of the lattice. They correspond to tunneling amplitudes between different lattice sites, except for the special case $\bm{i=0}$, $\nu=\nu'$ that corresponds to onsite energies.
Then, by defining
\begin{equation}
\hat{d}_{\nu{\bm{k}}}=\frac{1}{\sqrt{V_{B}}}
\sum_{\bm{j}} ~e^{-i{\bm{k}}\cdot{\bm{R}}_{\bm{j}}}\hat{a}_{\bm{j}{\nu}},
\label{eq:opk}
\end{equation}
${\hat{\cal{H}}}_0$ is transformed as
\begin{equation}
\hat{\cal{H}}_{0}=\sum_{\nu\nu'}\int_{\cal B} {d}^{2}{\bm{k}}~h_{\nu\nu'}({\bm{k}})
\hat{d}_{\nu{\bm{k}}}^{\dagger}\hat{d}_{\nu'\bm{k}}
\label{eq:intermh0}
\end{equation}
with
\begin{equation}
h_{\nu\nu'}({\bm{k}})=\sum_{\bm{i}}e^{i{\bm{k}}\cdot{\bm{R}}_{\bm{i}}}\langle {w}_{\bm{0}\nu}|\hat{H}_{0}|w_{\bm{i}\nu'}\rangle
\label{eq:hamnu}
\end{equation}
being the Hamiltonian density in quasi-momentum space, whose eigenvalues are in principle equal to the exact energy bands $\varepsilon_{\nu}(\bm{k})$. 
For practical purposes, however, the expression (\ref{eq:hamnu}) must be truncated 
by retaining only a finite number of matrix elements.
This corresponds to the tight-binding expansion in $\bm{k}$-space. The actual number of terms needed to reproduce the energy bands (or any other physical quantity) within a certain
degree of accuracy crucially depends on the properties of the basis functions $w_{\bm{j}\nu}(\bm{x})$, the MLWFs being the optimal choice due to their minimal spread, see Appendix \ref{app:a}.

From now on we will apply the above tight-binding expansion to the tunable two-dimensional honeycomb potential of the experiment \cite{tarruell2012}.

\section{Tunable honeycomb lattice}
\label{sec:honeycomb}

The functional form of the potential reproduced experimentally in Ref. \cite{tarruell2012} is
\begin{eqnarray}
\label{eq:tarruellpot}
&&V(x,y)=-V_{\overline{X}}\cos^2(k_{L}x+\theta/2)-V_{X}\cos^2(k_{L}x)\\
&&\quad-V_{Y}\cos^2(k_{L}y)-2\alpha\sqrt{V_{X}V_{Y}}\cos(k_{L}x)\cos(k_{L}y)\cos(\varphi)
\nonumber
\end{eqnarray}
where all the parameters can be controlled and tuned in the experiment. 
In particular, by varying the laser intensities 
$V_{\overline{X}},V_{X}$ and $V_{Y}$, several structures
can be realized by continuous deformations, ranging from chequerboard to triangular, dimer, honeycomb, 
and square lattices, including 1D chains.
\begin{figure}[t!]
\centerline{\includegraphics[width=0.9\columnwidth]{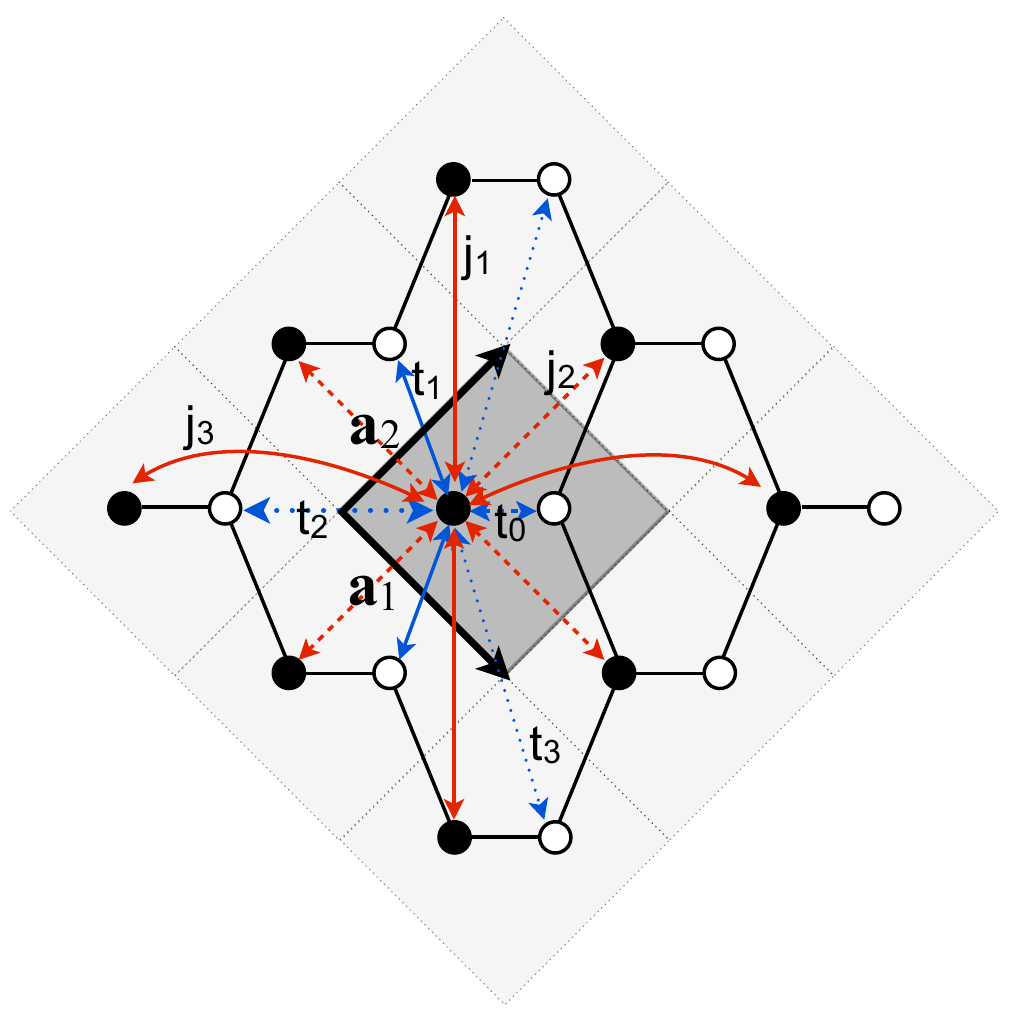}}
\caption{(Color online) 
Bravais lattice associated to the potential in Eq. (\ref{eq:tarruellpot})
for the stretched honeycomb
configuration. Black and white circles refers to minima of type $A$ and $B$, respectively. The elementary cell is highlighted in gray. The various diagonal and off-diagonal tunneling coefficients of our tight-binding expansion are indicated for the site of type A in the central cell.}
\label{fig:bravais}
\end{figure}
Let us define the Bravais lattice associated to the potential minima as
${\cal{B}}=\{\bm{R}_{mn}=m\bm{a}_{1}+n\bm{a}_{2}\Big| m,n\in \mathbb{Z}\}$ (see Fig. \ref{fig:bravais}) which is generated by the two basis vectors
\begin{equation}
\label{eq:basisvectors}
\bm{a}_{1,2}=\frac{\pi}{k_{L}}(\bm{e}_{x}\mp\bm{e}_{y}),
\end{equation}
with $\bm{e}_{x,y}$ being the cartesian unit vectors.
Therefore, the basis vectors in reciprocal space are
\begin{equation}
\bm{b}_{1,2}=k_{L}(\bm{e}_{x}\mp\bm{e}_{y}),
\end{equation}
following from $\bm{a}_{i}\cdot\bm{b}_{j}=2\pi\delta_{ij}$.
From now on, we can fix $k_{L}=1$, $\hbar=1$, $m=1/2$ without loss of generality. This corresponds to measure lengths in units of $1/k_{L}$ and energies in units of the recoil energy $E_{R}=\hbar^{2}k_{L}^{2}/2m$ \cite{ibanez2013}.

As the unit cell in direct space generally contains two basis points $A$ and $B$,
we consider the mixing of the two lowest bands.
It is then customary to write the Hamiltonian density in Eq. (\ref{eq:hamnu}) as
\begin{equation}
h(\bm{k})=\left(\begin{array}{cc}
 \epsilon_{A}(\bm{k}) & z(\bm{k}) \\
 z^{*}(\bm{k}) & \epsilon_{B}(\bm{k})
\end{array}\right)
\label{eq:hmatrix}
\end{equation}
where (see also Eq. (\ref{eq:psiexpf})) the band index $\nu=1,2$ has been traded to $\nu=A,B$ as the resulting
MLWFs are centered at subwells located in $A,B$.
The two lowest energy bands are then given by the eigenvalues of (\ref{eq:hmatrix})
\begin{equation}
\varepsilon_{\pm}(\bm{k})=\epsilon_{+}(\bm{k})\pm\sqrt{\epsilon_{-}^{2}(\bm{k})+|z(\bm{k})|^{2}}
\label{eq:eigenvaluesgen}
\end{equation}
with $\epsilon_{\pm}(\bm{k})=(\epsilon_{A}(\bm{k})\pm\epsilon_{B}(\bm{k}))/2$.

The matrix elements in (\ref{eq:hmatrix}) can be expanded as
\begin{eqnarray}
\epsilon_{\nu}(\bm{k})&=&\sum_{mn}J_{mn}^{\nu}e^{-i\bm{k}\cdot\bm{R}_{mn}}
\\
z(\bm{k})&=&-\sum_{mn}T_{mn}e^{-i\bm{k}\cdot\bm{R}_{mn}}
\end{eqnarray}
with
\begin{eqnarray}
J_{mn}^{\nu}&\equiv&\langle w_{\nu}^{\bm{0}}|\hat{H}_{0}|w_{\nu}^{\bm{R}_{mn}}\rangle\\
T_{mn}&\equiv&-\langle w_{A}^{\bm{0}}|\hat{H}_{0}|w_{B}^{\bm{R}_{mn}}\rangle
\label{eq:jandt}
\end{eqnarray}
corresponding to diagonal and off-diagonal matrix elements, respectively. 
The sign convention is chosen in such a way that all the coefficients appear positive defined \cite{ibanez2013}.
Here we truncate the tight-binding expansion by including all possible tunneling between neighboring cells, as indicated in Fig. \ref{fig:bravais} \cite{universal}.

Let us start by considering the diagonal terms. By fixing an arbitrary energy offset, we can write
\begin{eqnarray}
\label{eq:epsa}
\epsilon_{A}(\bm{k})&=&\epsilon + F^{A}({\bm{k}})\\
\epsilon_{B}(\bm{k})&=&-\epsilon + F^{B}({\bm{k}})
\label{eq:epsb}
\end{eqnarray}
with
\begin{equation}
\epsilon =(J^{A}_{00}-J^{B}_{00})/2
\label{eq:epsilon}
\end{equation}
and
\begin{eqnarray}
F^{\nu}(\bm{k})&=&2j^{\nu}_{1}\cos\left(2\pi k_{y}\right)+4j^{\nu}_{2}\cos\left(\pi k_{y}\right)\cos
\left(\pi k_{x}\right)\nonumber\\
&&+2j^{\nu}_{3}\cos\left(2\pi k_{x}\right).
\label{eq:effenu}
\end{eqnarray}
The tunneling coefficients appearing in Eq. (\ref{eq:effenu}) are precisely those
connecting the 
minima located at points of the same type $A$ or $B$ (see Fig. \ref{fig:bravais}), and have been redefined as follows
\begin{eqnarray}
j^{\nu}_{1}&\equiv&J^{\nu}_{1-1}=J^{\nu}_{-11}\nonumber\\
j^{\nu}_{2}&\equiv&J^{\nu}_{10}=J^{\nu}_{01}=J^{\nu}_{0-1}=J^{\nu}_{-10}\nonumber\\
j^{\nu}_{3}&\equiv&J^{\nu}_{11}=J^{\nu}_{-1-1}
\label{eq:tunel-coefs-j}
\end{eqnarray}
in order to simplify the notations. The form of the function $F^{\nu}(\bm{k})$ follows from the explicit form of the corresponding lattice vectors $\bm{R}_{mn}$.

We notice that when $\theta=\pi$ the minima in $A,B$
are degenerate in energy, so that $\epsilon=0$.
Also, $j^{A}_{i}=j^{B}_{i}\equiv j_{i}$, so that $F^{A}({\bm{k}})=F^{B}({\bm{k}})\equiv F({\bm{k}})$,
and the eigenvalues in Eq. (\ref{eq:eigenvaluesgen}) take the following simple form
\begin{equation}
\varepsilon_{\pm}(\bm{k})=F(\bm{k})\pm|z(\bm{k})|.
\label{eq:symbands}
\end{equation}

As far as the off-diagonal matrix element $z(\bm{k})$, its analytical form is
\begin{eqnarray}
z(\bm{k})&\equiv& -\left[t_{0}+2t_{1}\cos(\pi k_{y})e^{-i\pi k_{x}}+t_{2}e^{-2i\pi k_{x}}
\right.\nonumber\\
&&\left.+2t_{3}\cos(2\pi k_{y})\right]
\label{eq:zetatb}
\end{eqnarray}
where the tunneling coefficients have been redefined as
\begin{eqnarray}
t_{0}&\equiv&T_{00}\nonumber\\
t_{1}&\equiv&T_{10}=T_{01}\nonumber\\
t_{2}&\equiv&T_{-1-1}\nonumber\\
t_{3}&\equiv&T_{1-1}=T_{-11}.
\label{eq:tunel-coefs-t}
\end{eqnarray}
Notice that the ordering of the tunneling coefficients in Eqs. (\ref{eq:tunel-coefs-t}),(\ref{eq:tunel-coefs-j})
does not necessarily correspond to the hierarchy of their magnitudes, as this may depend on the regime of the potential parameters (see later on).

In the following,  we will use this model to discuss the features of the stretched honeycomb configuration \cite{note},  
which allows us to analyze the behavior of the Dirac points.
We will also discuss the effect of increasing the overall potential intensity (in order to enter a  well defined 
tight-binding regime, but with the same potential structure) and that of breaking the degeneracy between sites
of type $A$ and $B$.

\section{MLWF\lowercase{s} and tunneling coefficients for the degenerate case}

\begin{figure}[t!]
\centerline{\includegraphics[width=0.95\columnwidth]{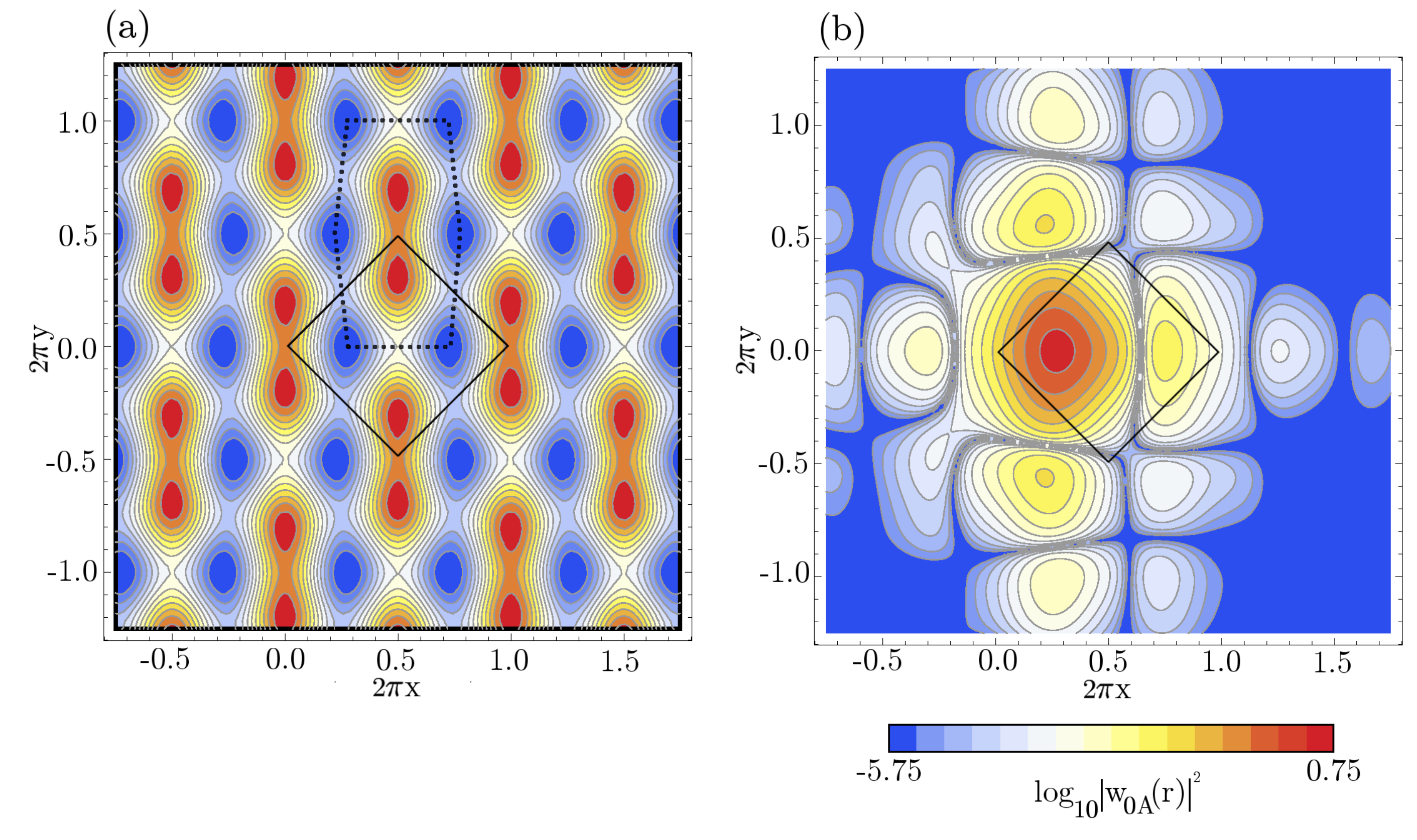}}
\caption{(Color online) (a) Stretched-honeycomb potential (\ref{eq:tarruellpot})  for $V_{\overline{X}}=5$ , $V_{X}=0.28 $, $V_{Y}=1.8 $. 
Hot and cold colors denote high and low values of the potential, respectively. 
The unit cell is indicated by solid (black) lines. 
(b) Structure of the calculated MLWF for sublattice $A$
for the same potential setup as in (a) (see text).}
\label{fig:sym-potential-mlwf}
\end{figure}

In this section we will discuss the numerical results for the stretched-honeycomb configuration with two degenerate minima per unit cell, obtained with  $\theta=\pi$, $\varphi=0$ in (\ref{eq:tarruellpot}). This is the most interesting configuration due to the presence of massless Dirac points (see later on).
The effect of parity breaking ($\theta\neq\pi$), that generates a Dirac mass,  will be analyzed in the next section. 
In addition, in Appendix \ref{app:b} we will present the results for a wider range of lattice configurations. 

\begin{figure}[b!]
\centerline{\includegraphics[width=0.82\columnwidth]{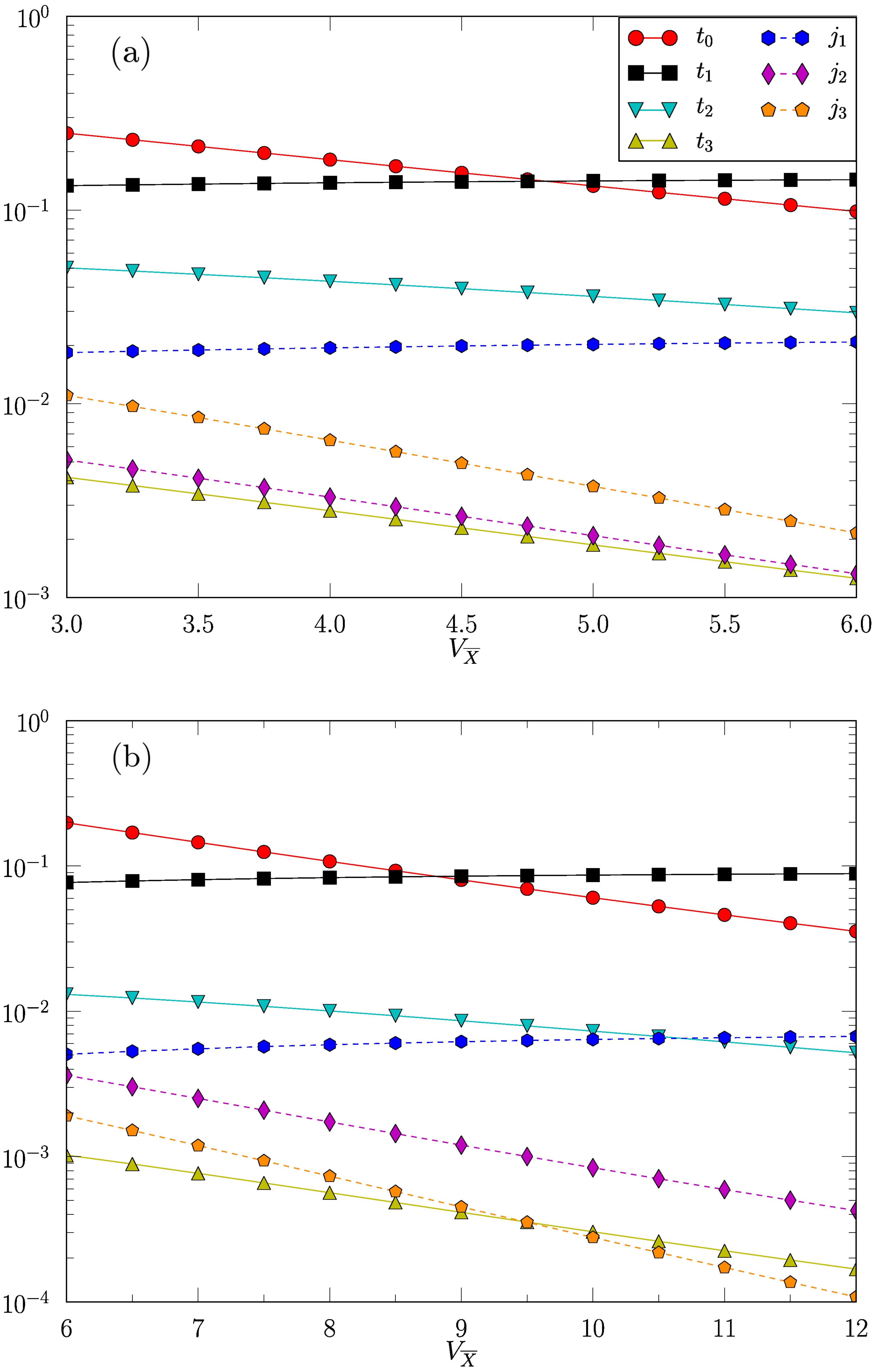}}
\caption{(Color online) Behavior of the various tunneling coefficients as function of $V_{\overline{X}}$.
(a) covers the experimental regime, $V_{X}=0.28 $, $V_{Y}=1.8 $, while
in (b) we consider a proper tight-binding regime, $V_{X}=0.56 $, $V_{Y}=3.6 $.
}
\label{fig:sym-tunnel-tarruel-str-hon}
\end{figure}

Here we start by considering the experimental 
regime of Tarruel \textit{et al.} \cite{tarruell2012}, namely
$V_{X}=0.28 $, $V_{Y}=1.8 $
and $V_{\overline{X}}$ variable ranging 
from $3$ to $6$. Within this range of parameters, the potential (\ref{eq:tarruellpot})
has the stretched-honeycomb structure shown in 
Fig. \ref{fig:sym-potential-mlwf}(a). 
This configuration determines the shape of the calculated MLWFs, drawn in Fig. \ref{fig:sym-potential-mlwf}(b)
for the sublattice of type $A$.
As shown in this figure, the MLWF is exponentially localized around the 
$A$ site of the central unit cell (note the logarithmic scale), 
and it presents a
non-negligible contribution around the 
neighboring potential minima, as well.
The associated tunneling coefficients are presented in Fig. \ref{fig:sym-tunnel-tarruel-str-hon}(a).
This figure shows the behavior of the diagonal and off-diagonal coefficients, $t_{i}$ ($i=0,3$) and $j_{i}$ ($i=1,3$) in Eqs. (\ref{eq:tunel-coefs-t}) and (\ref{eq:tunel-coefs-j}) respectively, as a function of $V_{\overline{X}}$.

\begin{figure}
\centerline{\includegraphics[width=0.9\columnwidth]{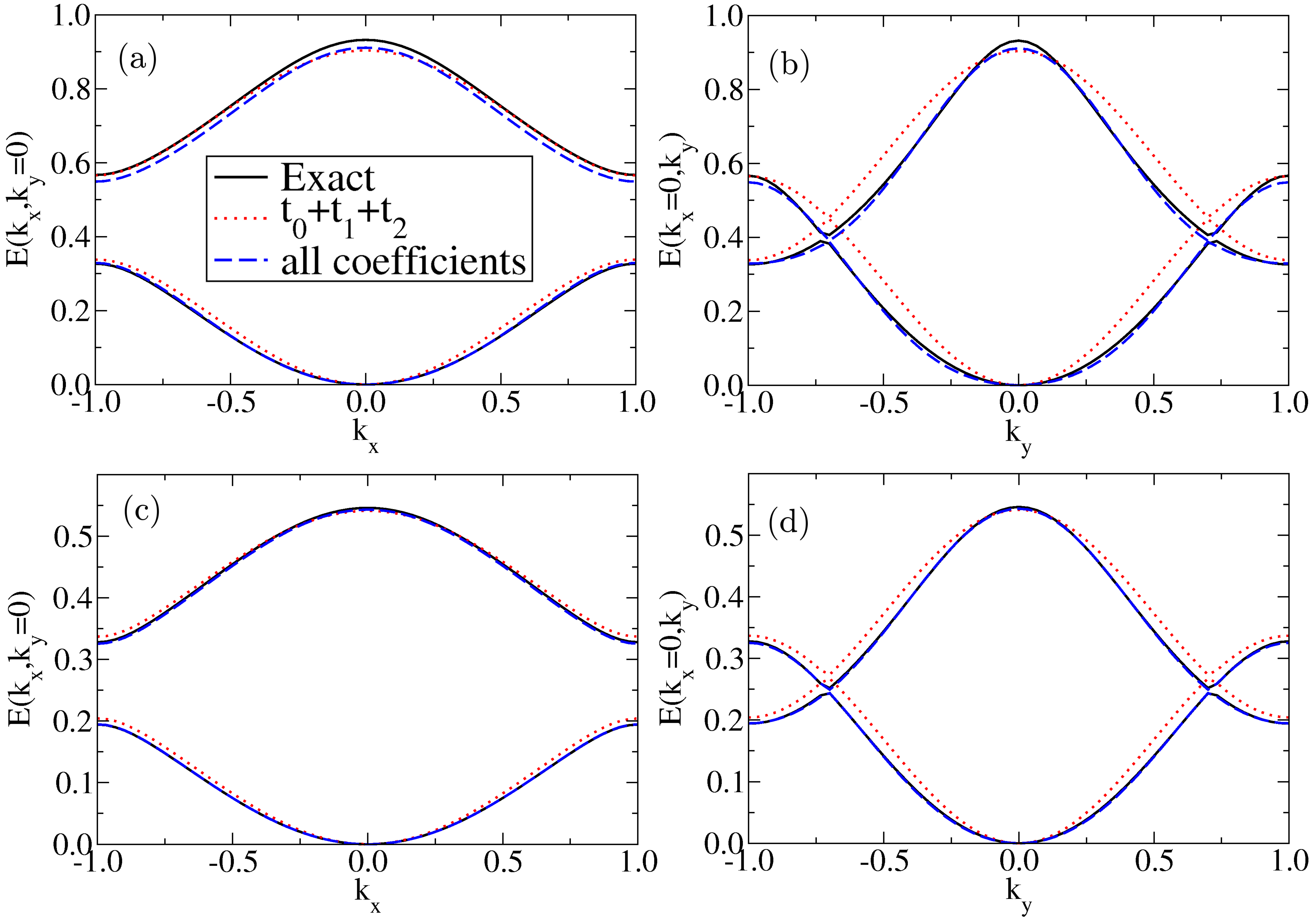}}
\caption{(Color online) 
Cut of the exact energy bands (black solid line) compared to the two tight-binding 
approximations with just $t_{0}$, $t_{1}$ and $t_{2}$ (red dotted line), and with all
the coefficients in Fig. \ref{fig:bravais} (green dotted-dashed line).
Panels (a) and (b) respectively show cuts along
$k_{x}$ ($k_{y}=0$) and $k_{y}$ ($k_{x}=0$), for 
$V_{X}=0.28 $, $V_{Y}=1.8 $ and $V_{\overline{X}}=5$;
(c) and (d) refer to
$V_{X}=0.56 $, $V_{Y}=3.6 $
and $V_{\overline{X}}=8.5$. 
}
\label{fig:sym-bands-apprx}
\end{figure}

With these parameters we then compute the tight-binding energy dispersion 
in Eq. (\ref{eq:symbands}), shown in Fig. \ref{fig:sym-bands-apprx} for some examples.
In particular, we  consider two different tight-binding approximations, one including just 
$t_{0}$, $t_{1}$ and $t_{2}$ (corresponding to the universal hamiltonian of Ref. \cite{montambaux2009}),
 and that including all the coefficients in Fig. \ref{fig:bravais}.
In Figs. \ref{fig:sym-bands-apprx}(a-b) we compare the energy dispersion of
the different tight-binding approximations with the exact spectrum
at $V_{\overline{X}}=5$.
The figures show that the main features, including the band-crossing along the $k_{y}$ direction 
(Fig. \ref{fig:sym-bands-apprx}(b)), are well reproduced by both approximations,
though the tight binding model with just $t_{0}$, $t_{1}$ and $t_{2}$ is not capable of approximating the exact bands with sufficient accuracy (this holds in all the range of $V_{\overline{X}}$ considered in this paper).
In any case, remarkably the model is able to reproduce with sufficient accuracy the position of the
Dirac points even in this parameter range, as shown in the next section.

Here we also consider a different set of values
for the potential parameters that correspond to a well defined
tight-binding regime, while maintaining the stretched-honeycomb structure.
In particular, we use the parameter values $V_{X}=0.56 $, $V_{Y}=3.6 $
and $V_{\overline{X}}$ ranging from $6$ to $12$, 
corresponding to twice the values of Tarruel \textit{et al.} \cite{tarruell2012}. 
The calculated tunneling coefficients are 
illustrated in Fig. \ref{fig:sym-tunnel-tarruel-str-hon}(b),
showing the same general structure as the ones in Fig. \ref{fig:sym-tunnel-tarruel-str-hon}(a),
except for minor differences regarding the smallest coefficients.
The corresponding energy dispersion is shown in Figs. \ref{fig:sym-bands-apprx}(c-d)
for $V_{\overline{X}}=8.5$ as an example. In this case, 
already the lowest order approximation with just the coefficients just 
$t_{0}$, $t_{1}$ and $t_{2}$ provides a remarkable agreement with the exact data.
This has been verified also for the other values of $V_{\overline{X}}$ in the range considered.

\section{Dirac points}

\begin{figure}[t!]
\centerline{\includegraphics[width=0.9\columnwidth]{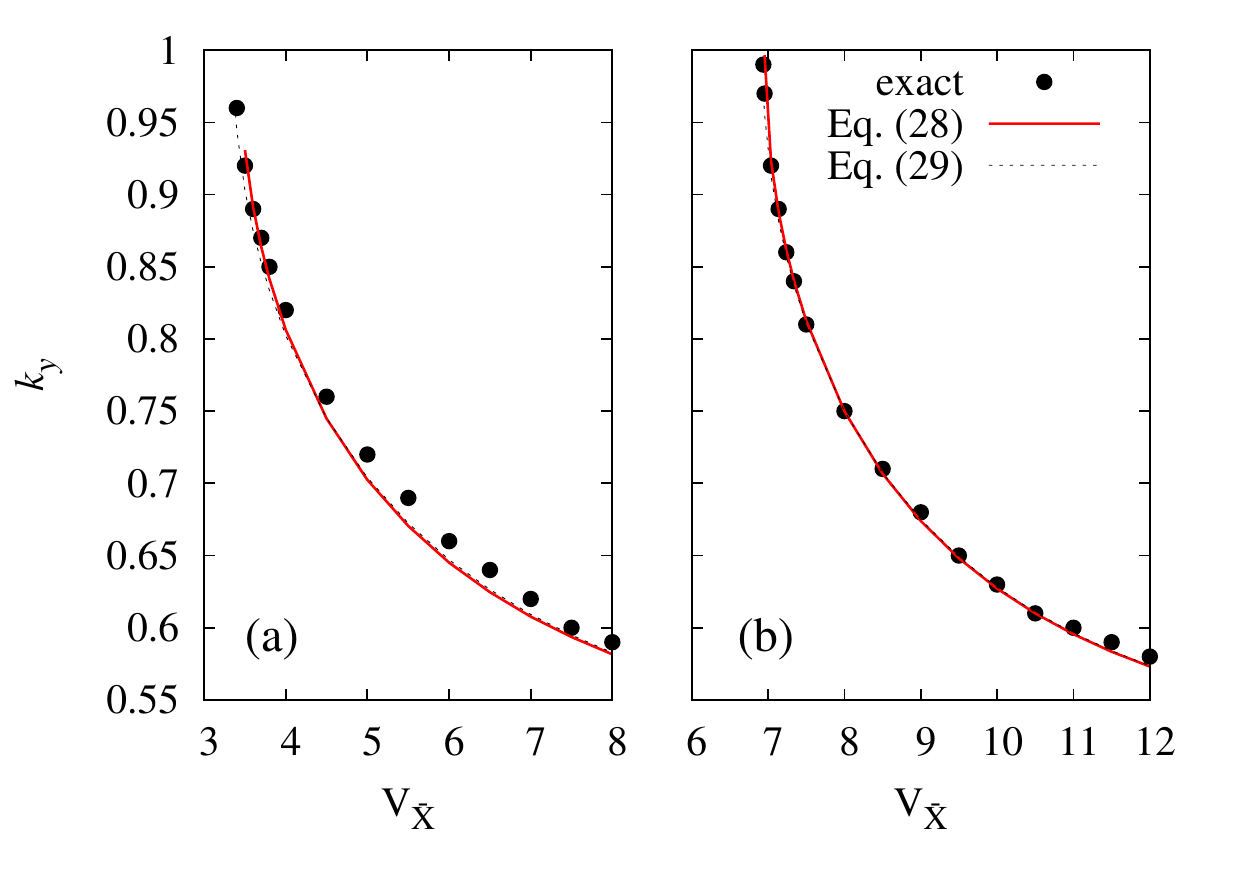}}
\caption{(Color online) Position of the Dirac points along the $k_{y}$-axis as a function of $V_{\overline{X}}$ for (a) the parameter regime of Tarruel \textit{et al.} \cite{tarruell2012}, and (b) the tight binding regime discussed in the text. 
The exact positions (circles) extracted from the Bloch spectrum are compared with the predictions of Eqs. (\ref{eq:diracpointloc}) and (\ref{eq:diracmont}).}
\label{fig:dirac}
\end{figure}

As we have seen in the previous section, the spectrum for a stretched honeycomb configuration with $\theta=\pi$ is characterized by points where the two bands are degenerate, with a linear dispersion along at least one direction - the so-called Dirac points. They are defined by $z(\bm{k}_{D})=0$ and come always in pairs due to time-reversal invariance, with $z^{*}(\bm{k}_{D})=z(-\bm{k}_{D})$ \cite{montambaux2009}. Their existence and position depends on the geometry of the lattice: for a regular honeycomb structure (a graphene-like lattice) they are located at the corners of the Brillouin zone \cite{lee2009,ibanez2013}, whereas in the present tunable case they can be moved inside the Brillouin zone, as showed in \cite{tarruell2012}. In particular, from the expression in Eq. (\ref{eq:zetatb}),
the position $\bm{k}_{D}=(k_{x},k_{y})$ of the Dirac points is obtained by solving the following equation
\begin{equation}
t_{0}+2t_{1}\cos(\pi k_{y})e^{-i\pi k_{x}}
+t_{2}e^{-2i\pi k_{x}}+2t_{3}\cos(2\pi k_{y})=0
\label{eq:diracpoints}
\end{equation}
whose imaginary part yields 
\begin{equation}
k_{x}=0
\end{equation}
inside the first Brillouin zone. Then, Eq. (\ref{eq:diracpoints}) becomes
\begin{equation}
t_{0}+2t_{1}{\rm{cos}}(\pi k_{y})+t_{2}+2t_{3}{\rm{cos}}(2\pi k_{y})
=0
\end{equation}
solved by
\begin{equation}
k_{y}=\pm \frac{1}{\pi}\cos^{-1}\!\!\left[\displaystyle{\frac{-t_{1}+
\sqrt{t_{1}^2+4t_{3}\left(2t_{3}-t_{0}-t_{2}\right)}}{4t_{3}}}\right].
\label{eq:diracpointloc}
\end{equation}
taking into account the considered hierarchy of the tunneling coefficients indicated in  Fig. \ref{fig:sym-tunnel-tarruel-str-hon}). 
In the current regimes, this expression can be further approximated as
\begin{equation}
 k_{y}\simeq \pm \frac{1}{\pi}\cos^{-1}\left[\displaystyle{-\frac{t_{0}+t_{2}}{2t_{1}}}\right]
\label{eq:diracmont}
\end{equation}
corresponding to the expression of Ref. \cite{montambaux2009}. Both Eq. (\ref{eq:diracpointloc}) and
its approximate version Eq. (\ref{eq:diracmont})  provide a valid solution when $t_{0}+t_{2}\leq 2t_{1}$, which is satisfied also in the range of parameters corresponding to the stretched honeycomb, as shown in Fig. \ref{fig:dirac}.
\begin{figure}
\centerline{\includegraphics[width=0.5\columnwidth]{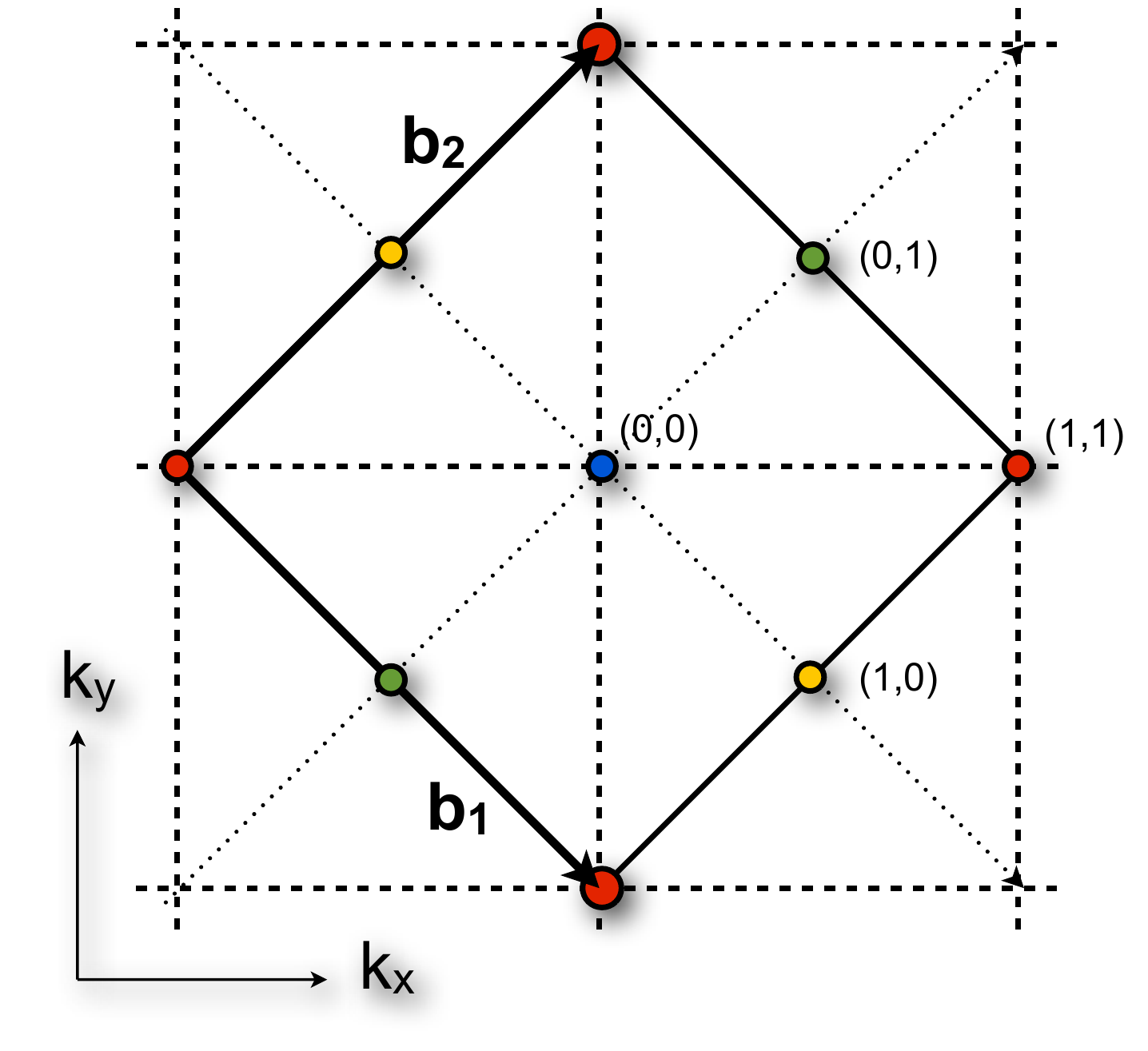}}
\caption{(Color online) Unit cell in quasimomentum space, with the location of the possible merging of the Dirac points. 
Equivalent points (connected by a reciprocal space vector $\bm{G}$)
are depicted with the same color. Given the actual values of the tunneling coefficients,
only the points at $k_{x}=0,k_{y}=\pm 1$ can be realized (larger red dots).
}
\label{fig:merging}
\end{figure}
\subsection{Merging of Dirac points}
\begin{figure*}[]
\centerline{\includegraphics[width=0.9\textwidth]{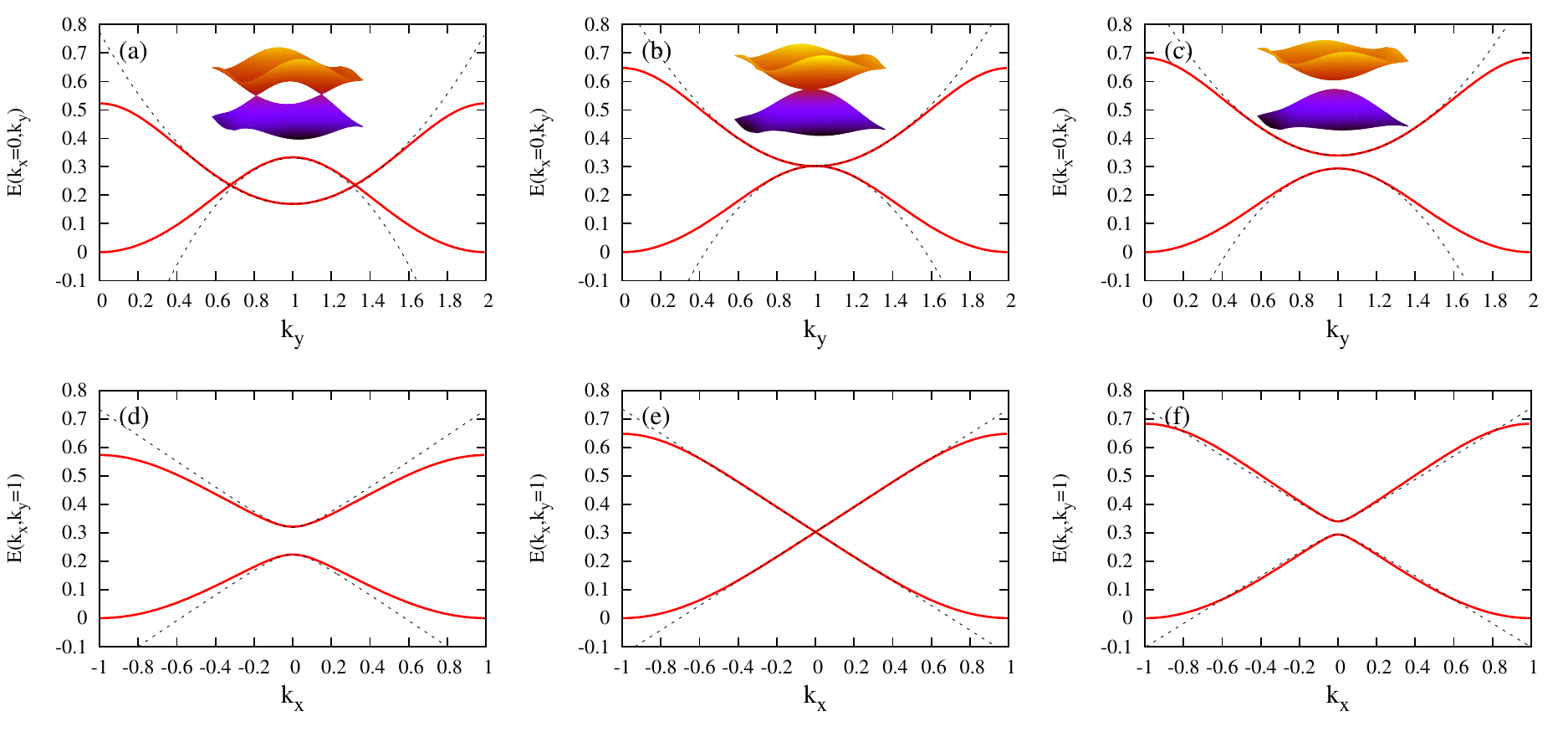}}
\caption{(Color online) Cuts of the energy bands around the merging point $\bm{k}_{M}= (0,1)$, for the tight binding regime discussed in the text ($V_{X}=0.56 $, $V_{Y}=3.6 $).
The exact Bloch bands (red solid lines) are compared to the approximate expressions in Eq. (\ref{eq:bandsclosemerging}), as a function of $k_{y}$ (at $k_{x}=0$) (a,b,c), and of $k_{x}$ (at $k_{y}=1$) 
(d,e,f). Each column correspond to a different value of $V_{\overline{X}}$: (a,d) $V_{\overline{X}}=8 $, (b,e) $V_{\overline{X}}=6.94 $ (merging point), $V_{\overline{X}}=6.54$ (c,f) . Note that the cut along $k_{x}$ in (d) does not cross the Dirac point, as the latter is located at $k_{y}\simeq 0.68$; this is the reason why Eq. (\ref{eq:bandsclosemerging}) provides a poorer approximation in this case.
}
\label{fig:merging-cut}
\end{figure*}
The merging of Dirac points occurs when the two solutions in Eq. (\ref{eq:diracpointloc}) coincide
modulo a reciprocal space vector $\bm{G}=p\bm{b}_{1}+q\bm{b}_{2}$ (with $p$, $q$ integers, see  Eq.
(\ref{eq:basisvectors})), namely at $\bm{k}_{M}=-\bm{k}_{M}+\bm{G}$.
Therefore, the merging point $\bm{k}_{M}$ satisfies \cite{montambaux2009}
\begin{equation}
\bm{k_{M}}=\bm{G}/2=\displaystyle{\frac{p\bm{b_{1}}+q\bm{b}_{2}}{2}}.
\end{equation}
In principle,  due to the geometry of the lattice, there are four possible
inequivalent merging points, at $(p,q)=(0,0), (0,1), (1,0), (1,1)$ \cite{montambaux2009}, see Fig. \ref{fig:merging}.
However, for the actual values of the tunneling coefficients,  
only the point $(1,1)$ and its equivalents are possible. 
In particular, in our case the Dirac points inside the first Brillouin zone merge (with those of outer cells) at the top and bottom corners $(1,-1)$ and $(-1,1)$, namely for $\bm{k}_{M}\equiv (0,\pm 1)$.
For the two examples considered here, see Fig. \ref{fig:dirac}(a) and (b), the merging occurs at $V_{\overline{X}}\simeq3.4$ (see also \cite{tarruell2012}) and $V_{\overline{X}}\simeq 6.94 $, respectively.

Following \cite{montambaux2009,lim2012}, we now expand the hamiltonian density around one of the two merging points,
defining $\tilde{\bm{k}}\equiv \bm{k}-\bm{k}_{M}$.
As discussed in Ref. \cite{lim2012}, the general form of the off-diagonal
component $z(\bm{k})$ around a merging point is characterized by a linear term in $\tilde{k}_{x}$ and a quadratic one in ${\tilde k}_y$, coming respectively from the imaginary and real parts of $z(\tilde{\bm{k}})$. Namely, the leading terms of the expansion are
\begin{eqnarray}
z_{R}(\tilde{\bm{k}})&\simeq&
-\left[t_{0}-2t_{1}+t_{2}+2t_{3}\right]+\pi^{2}\left[\left(4t_{3}-t_{1}\right)\tilde{k}_{y}^{2}\right]
\nonumber\\
z_{I}(\tilde{\bm{k}})&\simeq&2\pi\left(t_{2}-t_{1}\right)\tilde{k}_{x}.
\end{eqnarray}

Here we also take into account the diagonal term $F(\bm{k})$, not included in the approach of \cite{lim2012,montambaux2009}, as it affects the quadratic behavior introducing an asymmetry
between the two bands; neglecting an unimportant constant term we have
\begin{equation}
F(\tilde{\bm{k}})\simeq-2\pi^{2}(2j_{1}-j_{2})\tilde{k}_{y}^2.
\end{equation}

Therefore, close to the merging point the
hamiltonian  density can be cast into the form
\begin{equation}
h_{\nu\nu'}(\tilde{\bm{k}})\simeq \displaystyle{\frac{\tilde{k}_{y}^2}{2\mu}}\otimes I + \left(\Delta+\displaystyle{\frac{\tilde{k}_{y}^2}{2m^{*}}}\right)
\otimes\sigma_x + c\tilde{k}_{x}\otimes\sigma_y
\end{equation}
with
\begin{eqnarray}
\Delta&\equiv& -\left[t_{0}-2t_{1}+t_{2}+2t_{3}\right]
\label{eq:delta}\\
\displaystyle{\frac{1}{2m^{*}}}&\equiv&\pi^{2}\left(4t_{3}-t_{1}\right)\\
c&\equiv& 2\pi\left(t_{1}-t_{2}\right)\\
\displaystyle{\frac{1}{2\mu}}&\equiv& -2\pi^{2}(2j_{1}-j_{2}).
\label{eq:mu}
\end{eqnarray}
The corresponding dispersion law is
\begin{equation}
 \varepsilon_{\pm}(\tilde{\bm{k}})\simeq\displaystyle{\frac{\tilde{k}_{y}^2}{2\mu}}\pm\sqrt{\left(\Delta+\displaystyle{\frac{\tilde{k}_{y}^2}{2m^{*}}}\right)^{2}
+c^{2}\tilde{k}_{x}^2}
\label{eq:bandsclosemerging}
\end{equation}
with $\Delta$ being vanishing at the merging point, marking the topological transition between semi-metallic and  insulating phases driven by a change of sign in the product $m^{*}\Delta$ \cite{montambaux2009,lim2012}. The expression (\ref{eq:bandsclosemerging}) provides indeed a good approximation 
of the exact Bloch energies close to the merging point, as shown in Fig. \ref{fig:merging-cut} (a similar expansion can be derived around a generic Dirac point). In this picture we consider $\bm{k}_{M}=(0,1)$, that is the top corner of the first Brillouin zone in Fig. \ref{fig:merging}, and show band-cuts along orthogonal directions at $k_{x}=0$ (upper panels)
and $k_{y}=1$ (lower panels). 
Panels (a),(c) show two Dirac points belonging to adjacent Brillouin zones, symmetric with respect to $k_{y}=1$
(see Eq. (\ref{eq:diracpointloc})). This corresponds to a positive $\Delta$ 
(in the present regime of parameters $m^{*}$ is always negative). 
By decreasing $V_{\overline{X}}$ the two Dirac points approach each other and eventually merge 
 when $\Delta=0$, see panels (b,e). At this particular point and the dispersion law is linear
along $k_{x}$ and quadratic along $k_{y}$. 
By further decreasing $V_{\overline{X}}$, a gap opens at the merging point, see panels (c),(f).  In this case
the mass-like term is characterized by a negative $\Delta$. In all the panels, the
bands are compared with the low-energy expansion (34), showing a fine agreement close to the merging point
(except panel (d), see caption). 
In particular, a small asymmetry in the quadratic behavior along $k_{y}$  
is visible and well reproduced close to the merging point,
 owing to the diagonal term proportional to $1/2\mu$ in Eq. (\ref{eq:mu}) \cite{muterm}.

\subsection{Breaking parity: massive Dirac points}

\begin{figure}[]
\centerline{\includegraphics[width=\columnwidth]{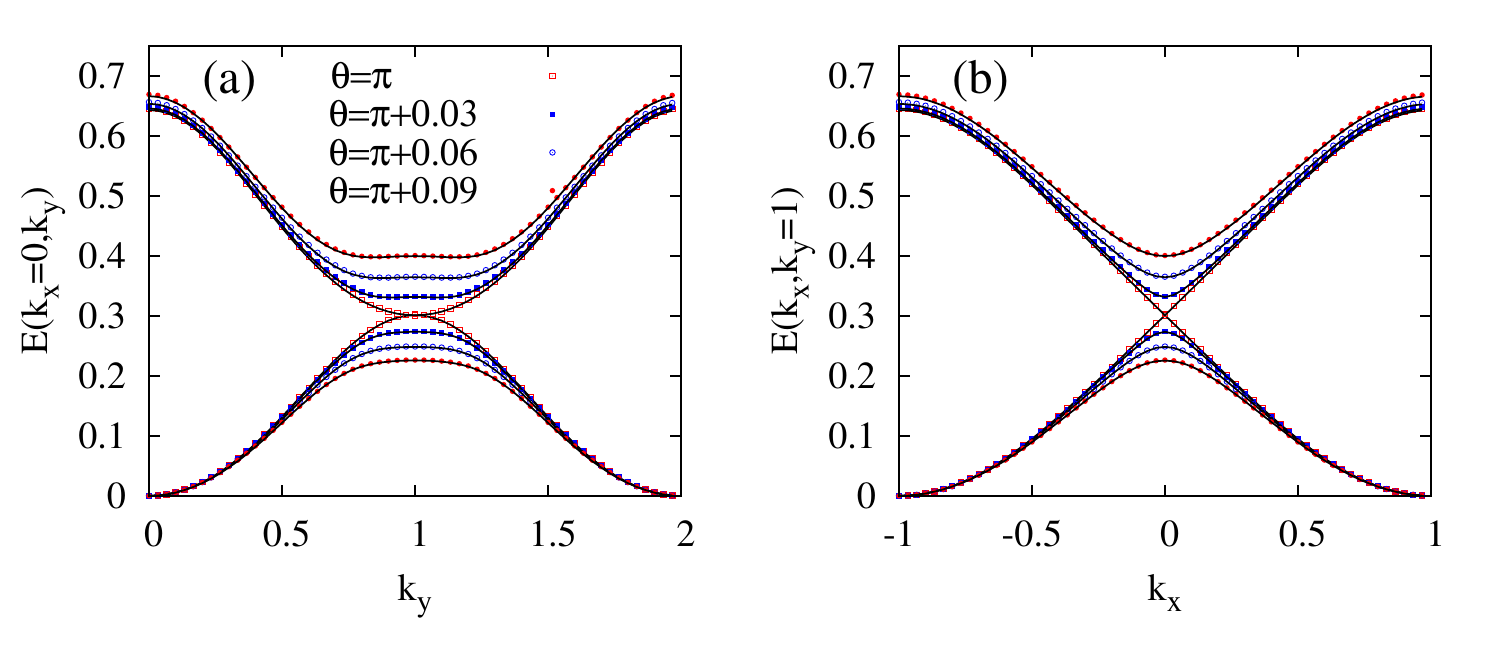}}
\caption{(Color online) 
Cuts of the energy bands around the merging point $\bm{k}_{M}= (0,1)$ for different values of the parity breaking angle $\theta$.
The exact Bloch bands (dots) are compared to the full tight-binding model (solid line), as a function of $k_{y}$ (at $k_{x}=0$) (a), and of $k_{x}$ (at $k_{y}=1$) (b).
The picture refers to the tight binding regime $V_{X}=0.56$, $V_{Y}=3.6$, $V_{\overline{X}}=6.54$ .
}
\label{fig:asym-pi}
\end{figure}

\begin{figure}[]
\centerline{\includegraphics[width=\columnwidth]{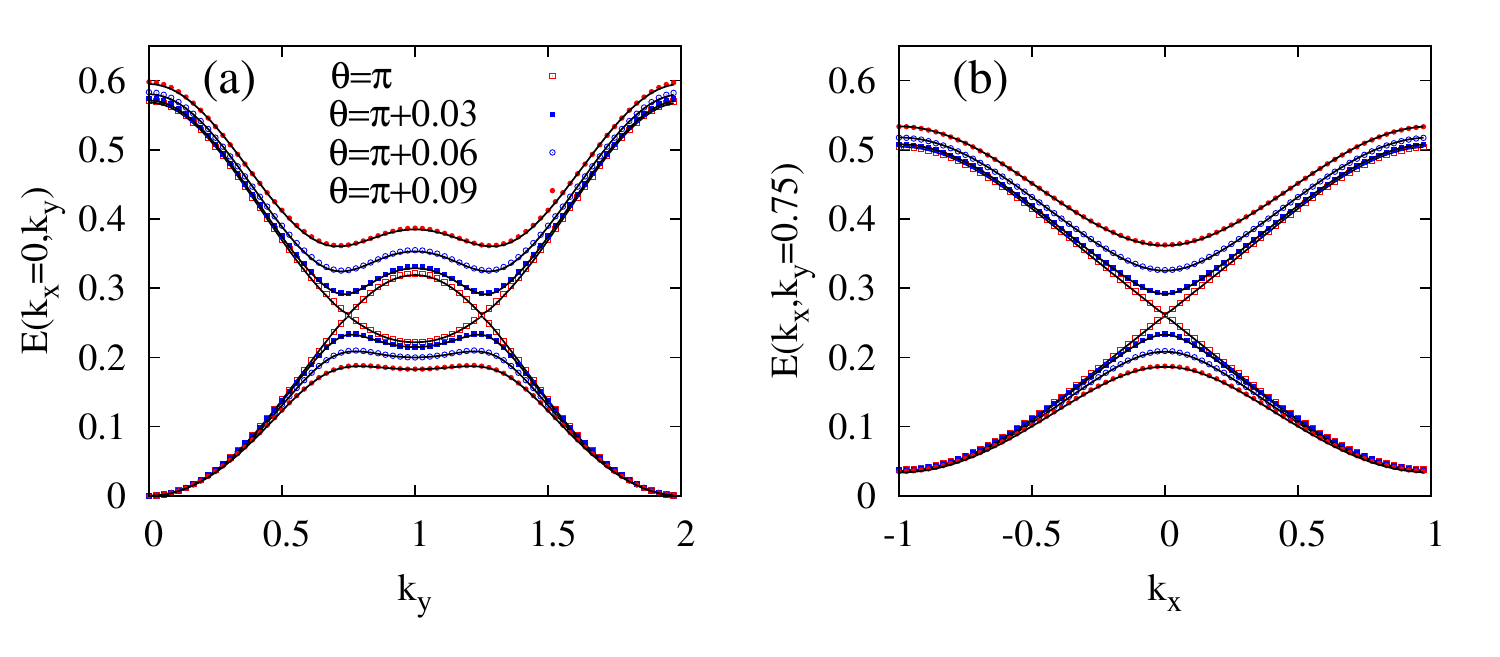}}
\caption{(Color online) Same as Fig. \ref{fig:asym-pi}, but for a case with two Dirac points located at $\bm{k}=(0,0.75)$ ($V_{X}=0.56$, $V_{Y}=3.6$, $V_{\overline{X}}=8.05$).}
\label{fig:asym-075}
\end{figure}

As shown in the experiment \cite{tarruell2012}, 
a gap can be opened at the Dirac points by breaking the invariance under parity, achieved by tuning the angle $\theta$ away from $\pi$. In this case, due to the asymmetry of two minima in the unit cell,  the diagonal terms $\epsilon_{\nu}$ and $j^{\nu}$ ($\nu=A,B$) are no longer degenerate, see Appendix \ref{app:d}. This causes the Dirac particles to acquire a mass, as it is evident from Figs. \ref{fig:asym-pi},\ref{fig:asym-075} where we show the energy bands for two  Dirac points, at $\bm{k}=(0,0.75)$ and at the merging point $\bm{k}_{M}=(0,1)$. 
This figures show that even small deviations from 
$\theta=\pi$ give rise to a significant mass term (gap) at the Dirac points, and that the the current tight-binding model accurately reproduce the exact energy bands. Note that in this case
the full formula (\ref{eq:eigenvaluesgen}) has to be used. We also mention that even in this case one can derive an expansion analog to that in Eq. (\ref{eq:bandsclosemerging}). 

\section{Conclusions}

Maximally localized Wannier functions for composite bands \cite{marzari1997} are a powerful tool 
for constructing tight binding models for ultra cold atoms in optical lattices.
Here, we have considered the tunable honeycomb optical lattice of Ref. \cite{tarruell2012}, and 
we have shown how to derive the corresponding tight-binding hamiltonian, \textit{ab initio}.  
We have calculated the MLWFs and the tunneling coefficients for different lattice configurations, 
showing that the spectrum properties, including the position of Dirac points and the dispersion law close to their merging, can be reproduced with high accuracy with an expansion up to third-nearest neighbors.
We have considered both cases of massless and massive Dirac points, respectively for the case of two degenerate minima per unit cell and for the case of parity breaking.
These results provide a direct connection between the experimental results of Ref. \cite{tarruell2012} and the universal hamiltonian of Refs. \cite{montambaux2009,lim2012}.

\acknowledgments
This work has been supported by the UPV/EHU under programs UFI 11/55 and IT-366-07, the Spanish Ministry of Science and Innovation through Grants No. FIS2010-19609-C02-00 and No. FIS2012-36673-C03-03, and the Basque Government through Grant No. IT-472-10.

\appendix
\section{Approximate Bloch spectrum}
\label{app:a}

Here we briefly review and comment about the use of the MLWFs for the tight binding expansion of the exact Bloch spectrum, following the discussion in Ref. \cite{marzari1997} (see also \cite{marzari2012}).
Let us start by rewriting the the hamiltonian density (\ref{eq:hamnu}) as follows (see Eq. (\ref{eq:mlwfs}))
\begin{eqnarray}
\label{eq:hgen}
&& h_{\nu\nu'}(\bm{k})=\sum_{\bm{i}}e^{i{\bm{k}}\cdot{\bm{R}}_{\bm{i}}}\langle {w}_{\bm{0}\nu}|\hat{H}_{0}|w_{\bm{i}\nu'}\rangle\\
&&=\frac{1}{V_{B}}\sum_{\bm{i}}\int_{\cal{B}}d\bm{q}~
e^{i(\bm{k-q})\cdot\bm{R}_{\bm{i}}}
\sum_{n}U^{*}_{\nu n}(\bm{q})U_{\nu'n}(\bm{q})\varepsilon_{n}(\bm{q})\nonumber
\end{eqnarray}
where $\varepsilon_{n}(\bm{k})$ are the (exact) Bloch bands, and the $U_{\nu n}(\bm{q})$ are periodic, unitary matrices representing \textit{gauge transformations} (as a function of quasimomentum) of the Bloch states.

\textit{Single bands.}
For $U(1)$ transformations that do not mix the bands, namely when 
\begin{equation}
U_{\nu m}(\bm{k})=e^{i\phi_{\nu}(\bm{k})}\delta_{\nu m} 
\end{equation}
the onsite energies and tunneling coefficients $\langle {w}_{\bm{0}\nu}|\hat{H}_{0}|w_{\bm{i}\nu'}\rangle$ in (\ref{eq:hgen}) are independent on
the (periodic) phases $\phi_{\nu}(\bm{k})$.
Furthermore, if all the terms in the sum are retained, owing to the following formula valid for an infinite lattice
\begin{equation}
\frac{1}{V_{B}}\sum_{\bm{i}}e^{i{\bm{R}}_{{\bm{i}}}\cdot (\bm{k}'-\bm{k})}=\delta(\bm{k}'-\bm{k})
\label{eq:deltak}
\end{equation}
one easily recovers the exact diagonal expression $h_{\nu\nu'}(\bm{k})=\delta_{\nu\nu'}\varepsilon_{\nu}(\bm{k})$.
This result is trivial, following directly from the completeness of the Wannier basis. 
As a consequence, the tunneling coefficients can be expanded in terms of the exact energies (with no reference to the Wannier functions) \cite{he2001,salerno2002}. In addition, the tight binding approximation of the exact Bloch spectrum, namely the truncation of the sum in (\ref{eq:hgen}) at a given order, is independent on the choice of the Wannier states. So, in the absence of band mixing (the gauge group being a direct product of $U(1)$ groups)
the tight binding expansion is \textit{gauge independent}. 

\textit{Composite bands.}
Let us now consider the case of composing $N$ bands, via a non-abelian $U(N)$ gauge transformation.
Again, summing over all lattice sites the hamiltonian density in Eq. (\ref{eq:hgen}) takes the form
\begin{equation}
 h_{\nu\nu'}(\bm{k})=\sum_{n}U^{*}_{\nu n}(\bm{k})U_{\nu'n}(\bm{k})\varepsilon_{n}(\bm{k})
\end{equation}
whose eigenvalues coincide with the exact bands $\varepsilon_{\nu}(\bm{k})$ owing to the unitarity of the transformation. Moreover, even the trace of $h_{\nu\nu'}(\bm{k})$ in Eq. (\ref{eq:hgen}) 
at a finite order of the tight binding expansion is \textit{gauge-independent}, as
it does not dependent on the transformation matrices. Instead, finite order approximations of individual Bloch bands (or the sum of a subset of them) are \textit{gauge-dependent}, 
as they depend on a particular choice of the matrices $U_{nm}(\bm{k})$.

\textit{Parallel transport gauge.} We recall that the transformation matrices $U_{nm}(\bm{k})$ are defined as those that minimize the Wannier spread  
$\Omega=\sum_{\nu}[\langle \bm{r}^2\rangle_{\nu}-\langle \bm{r}\rangle_{\nu}^{2}]$, and that the latter can be decomposed as $\Omega=\Omega_I+\tilde{\Omega}$ \cite{marzari1997}. The first term is 
gauge invariant, while -- in case of composite bands  -- the second can be written as the sum of two (non negative) diagonal and off-diagonal components, $\tilde{\Omega}=\Omega_D+\Omega_{OD}$.
Both $\Omega_{D}$ and $\Omega_{OD}$ can be written in terms of the generalized Berry vector potentials $\bm{A}_{\nu\nu'}(\bm{k})$, 
defined as \cite{niu1995,pettini2011}
\begin{equation}
\bm{A}_{\nu\nu'}(\bm{k})=iV_{B}\langle u_{\nu\bm{k}}|\bm{\nabla}_{\bm{k}}|u_{\nu'\bm{k}}\rangle
\label{eq:connections}
\end{equation}
with the matrix $\bm{A}_{\nu\nu'}(\bm{k})$ being hermitian.
In one-dimension (1D), the gauge in which $\Omega$ is minimized 
corresponds to $\Omega_{OD}=0$ as a consequence of the vanishing of off-diagonal ($\nu\neq\nu'$) Berry connections
in Eq. (\ref{eq:connections}), and is called the \textit{parallel transport gauge}.
In this case, the transformation $U(k)$ can be obtained directly by requiring the off-diagonal connections to be vanishing \cite{pettini2011}, so that the optimal tight binding expansion can be seen as a direct property of the Hilbert space. Nevertheless, as the Wannier functions depend on the choice of the gauge, the two points of view are correlated. We remark that this approach is generally limited to 1D cases as in higher dimensions it is not always possible to make $\Omega_{OD}$ vanishing \cite{marzari1997}, so that the parallel-transport formulation can not be easily generalized. 
However, though in absence of a formal proof, in general we may assume that the gauge where the spread of Wannier functions is minimal, is the one that provides the best tight binding approximation of individual Bloch bands.
In fact, this has been already verified in a number of models \cite{modugno2012,ibanez2013,cotugno2013}.

Finally, we remark that the use of \textit{composite} instead of \textit{single band} transformations is required in case of a set of almost degenerate bands (well separated from the others), that usually corresponds to more that one minimum per unit cell, as in the present case. A more thorough discussion on this point, for the case of a 1D double well potential, can be found in \cite{modugno2012}.

\section{MLWF\lowercase{s} and tunneling coefficients}
\label{app:b}

\begin{figure}[b!]
\centerline{\includegraphics[width=0.8\columnwidth]{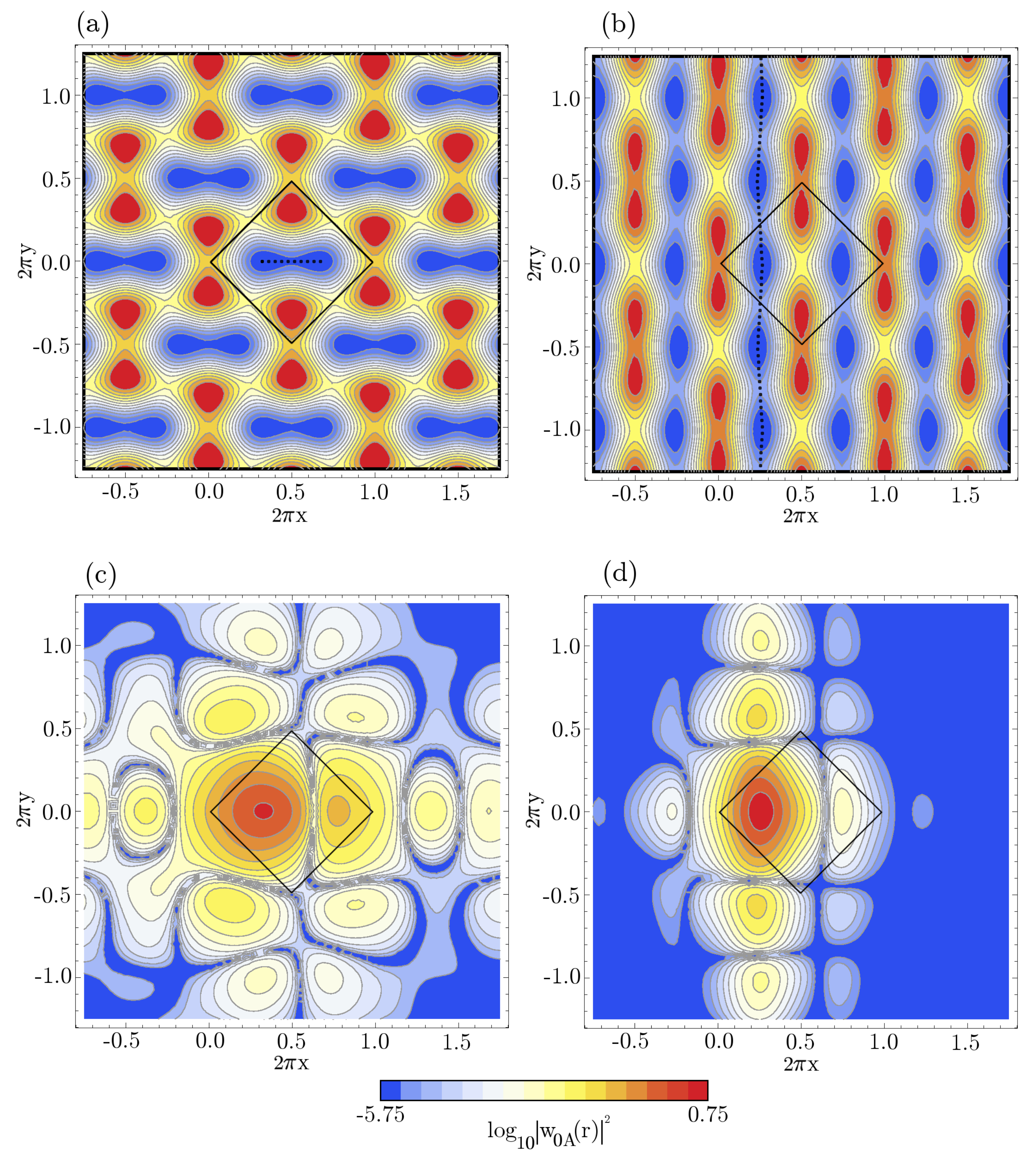}}
\caption{(Color online) Dimmer ($V_{\overline{X}}=1$) 
and 1D-chain ($V_{\overline{X}}=8$) limits for
fixed $V_{X}=0.28 $, $V_{Y}=1.8 $.
(a) and (b) show the respective potential structures, 
and (c) and (d) show associated MLWFs.
Color code as in Fig. \ref{fig:sym-potential-mlwf}.}
\label{fig:sym-potential-mlwf-dimer}
\end{figure}

In this appendix we analyze the properties of the  
MLWFs and the associated tunneling  coefficients in 
a range of $V_{\overline{X}}$ broader
than the one covered in the text.
This allows us to analyze the two opposite limits of the potential (\ref{eq:tarruellpot})
corresponding to the dimmer and 1D chain structures \cite{tarruell2012}.
These are exemplified in Figs. \ref{fig:sym-potential-mlwf-dimer}(a),(b) ($V_{\overline{X}}=1$ and $V_{\overline{X}}=8$, respectively) for the
experimental regime of Ref. \cite{tarruell2012}.
The dimmer structure is characterized by a 
relatively low value of the potential in the region between
 $A$ and $B$ within the same unit cell. 
On the opposite, in the 1D chain regime 
the potential is low along the $y$ direction connecting different minima, 
while it presents a barrier between the $A$ and $B$ sites of the same unit cell.
The stretched-honeycomb regime covered in the text
(Fig. \ref{fig:sym-potential-mlwf}) represents an intermediate configuration
between these two limits.

The structure of the potential in these limits 
determines the shape of
the MLWFs, which we illustrate in Figs. \ref{fig:sym-potential-mlwf-dimer}(c-d)
(results shown for sublattice $A$).
As in the stretched-honeycomb structure,
the MLWFs are exponentially localized around the 
$A$ site of the central unit cell and present a
non-negligible contribution around the 
neighboring potential minima.
In the case $V_{\overline{X}}=1$, see Fig. \ref{fig:sym-potential-mlwf-dimer}(c),
we find a large contribution of the MLWF 
around the $B$ site of the central unit cell, 
consistent with the dimmer structure of the potential. 
The situation is very different for $V_{\overline{X}}=8$, in
Fig. \ref{fig:sym-potential-mlwf-dimer}(d),
which shows a MLWF highly localized along the $y$ axis,
resembling the 1D chain structure of the potential.

\begin{figure}
\centerline{\includegraphics[width=0.7\columnwidth]{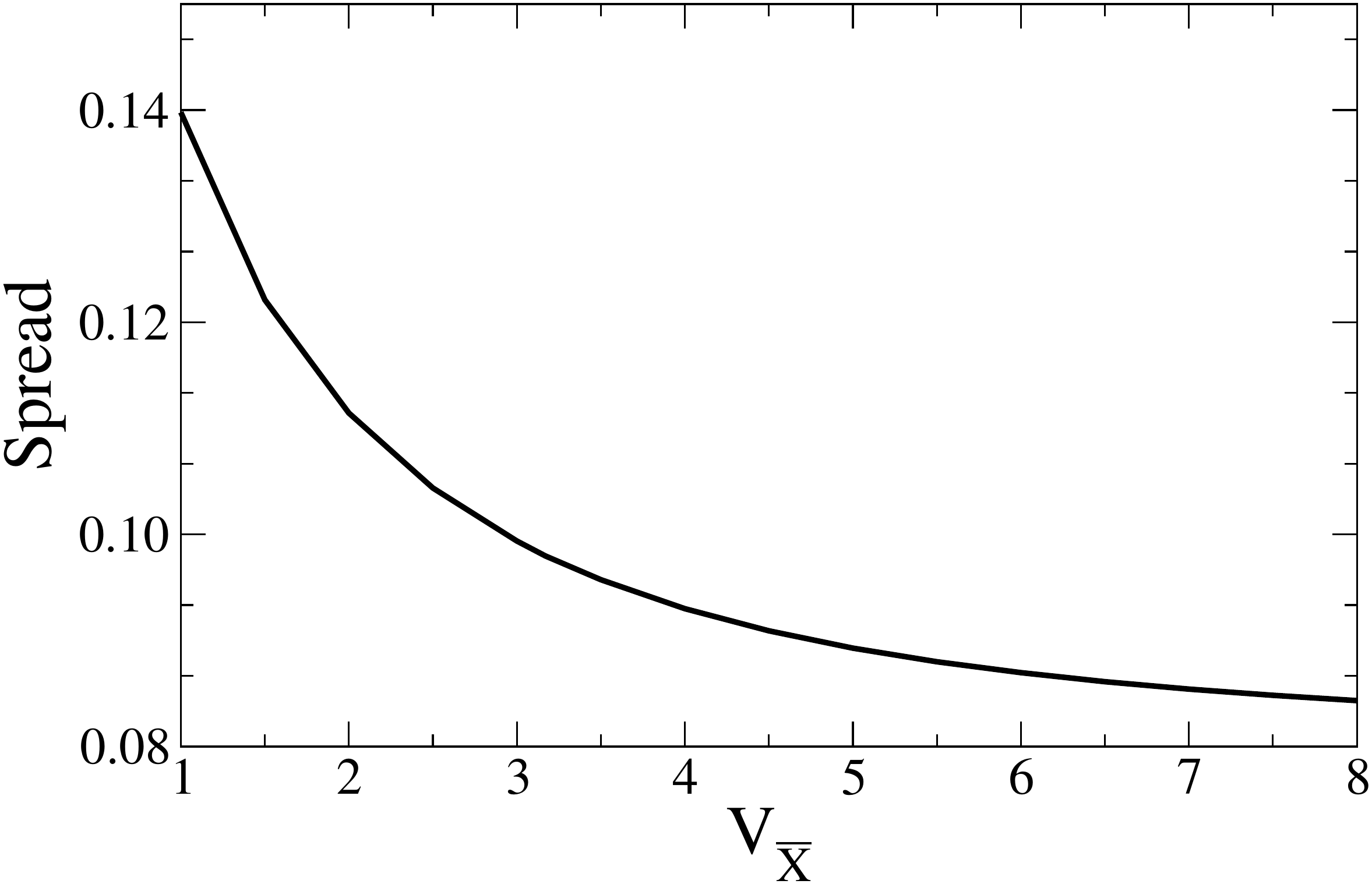}}
\caption{Spread of the MLWFs as a function of $V_{\overline{X}}$, 
in the regime of the experiment \cite{tarruell2012} ($V_{X}=0.28$, $V_{Y}=1.8$).
}
\label{fig:spread}
\end{figure}

In order to analyze 
the degree of localization of the MLWFs,
in Fig. \ref{fig:spread} we show the spread of the MLWFs, $\Omega=\sum_{\nu=1}^{2}[\langle \bm{r}^2\rangle_{\nu}-\langle \bm{r}\rangle_{\nu}^{2}]$
\cite{marzari1997}, as a function of $V_{\overline{X}}$.
The figure shows that, by increasing $V_{\overline{X}}$, $\Omega$ rapidly decreases in the regime of low $V_{\overline{X}}$,
while it almost saturates in the opposite limit. This indicates that
the tight-binding approach is expected to work better in the stretched-honeycomb  and the 1D-chain regimes, rather than for the dimmer case. 

\begin{figure}[t!]
\centerline{\includegraphics[width=0.8\columnwidth]{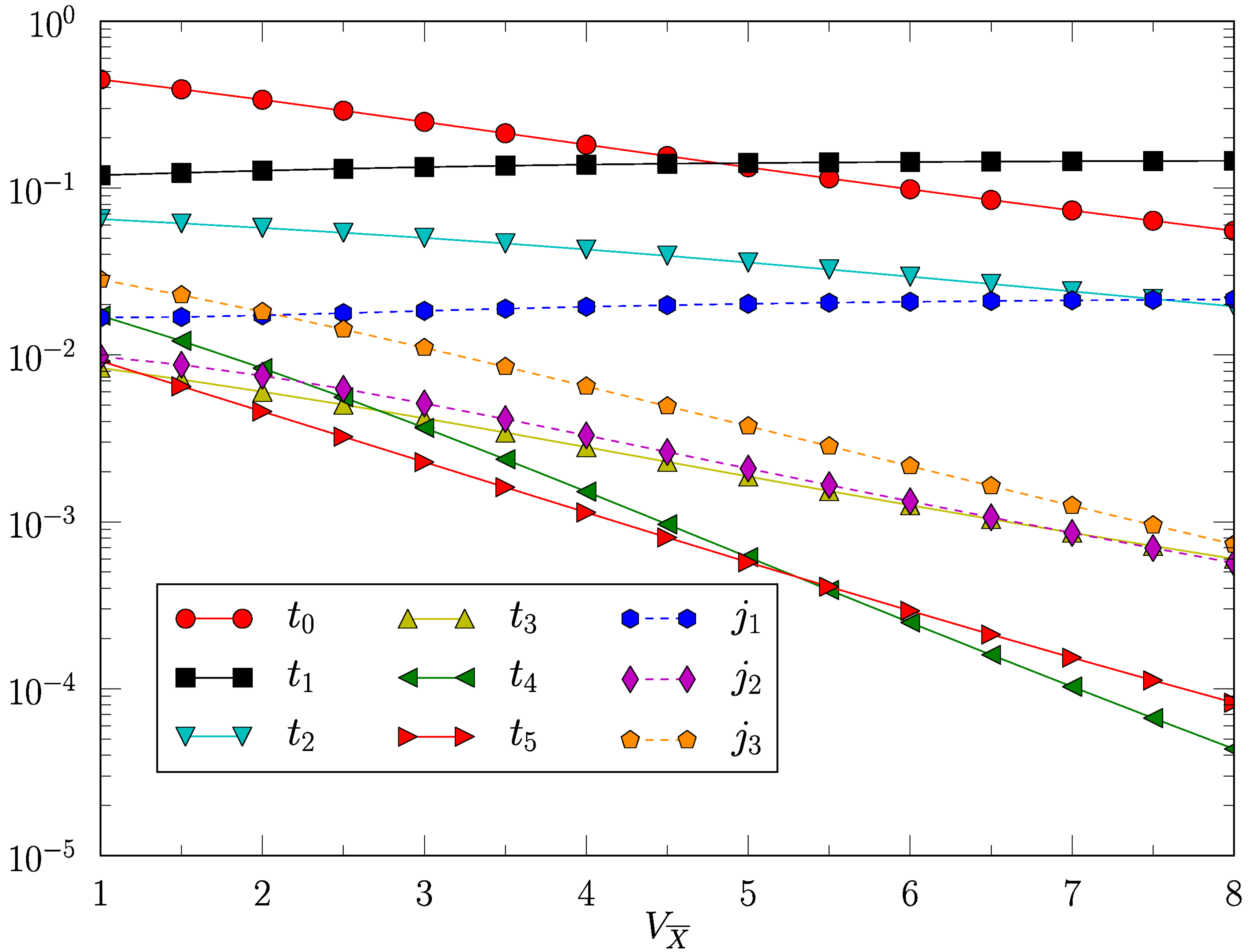}}
\caption{(Color online) Evolution of various tunneling coefficients as a function of $V_{\overline{X}}$,
covering the whole range from the dimmer ($V_{\overline{X}}\approx1$) to the 1D-chain  ($V_{\overline{X}}\approx1$) limits in the regime of the experiment \cite{tarruell2012} ($V_{X}=0.28$, $V_{Y}=1.8$).
}
\label{fig:sym-tunnel-tarruel}
\end{figure}

The behavior of the tunneling coefficients in the whole range from the
dimmer to the 1D chain limits are shown in Fig. \ref{fig:sym-tunnel-tarruel}. 
We first focus on the left hand side of the graphic, $V_{\overline{X}}\simeq 1E_{R}$.
There, we find that the ratio between the two dominant coefficients is $t_{0}/t_{1}\backsimeq10$.
This reflects the dimmer structure of the potential, 
since $t_{0}$ connects sites $A$ and $B$ (see Fig. \ref{fig:bravais}). 
Noteworthy, $t_{2}$ is by far the next biggest coefficient,
comparable  in magnitude to $t_{1}$. 
This reveals that the tunneling between
neighboring dimmers in $x$ direction is considerable (see $t_{2}$ in Fig. \ref{fig:bravais}).
The rest of the coefficients have a significantly lower value than $t_{0}$, $t_{1}$ and $t_{2}$.
We note that in Fig. \ref{fig:sym-tunnel-tarruel} there are two coefficients, 
$t_{4}\equiv T_{11}$ and $t_{5}\equiv -T_{-10}=-T_{0-1}$ (see Eq. \ref{eq:jandt}), 
that were
not considered in our expansion. 
In the dimmer regime, these coefficients can be 
larger than $j_{1}$, $j_{2}$ and $t_{3}$,
included in our tight-binding model. 

As $V_{\overline{X}}$ is increased, the various tunneling coefficients 
evolve in two different ways. Most of them decrease in magnitude, 
reflecting the stronger localization of the MLWFs as we approach a
more tight-binding regime. This could be termed the `normal' behavior, 
followed for instance by the tunneling 
coefficients of a perfect honeycomb lattice \cite{ibanez2013,note}.
However, two coefficients, namely $t_{1}$ and $j_{1}$,
increase in magnitude as $V_{\overline{X}}$ is increased. 
This `inverse' behavior reflects the evolution of the
potential (\ref{eq:tarruellpot}) from the dimmer to the 1D chain
structure,  as
these coefficients
connect potential minima inside the 1D chains. 
Owing to this `inverse' behavior, $t_{1}$ becomes the dominant coefficient for
$V_{\overline{X}}\gtrsim4.5E_{R}$. 
Similarly, $j_{1}$ becomes larger than $j_{3}$ and even $t_{2}$ 
for $V_{\overline{X}}\gtrsim7.5$. 
Thus, it is clear that varying the potential amplitude
can modify the role of the different tunneling coefficients.

\section{Accuracy of the tight-binding models}

\begin{figure}[]
\centerline{\includegraphics[width=0.9\columnwidth]{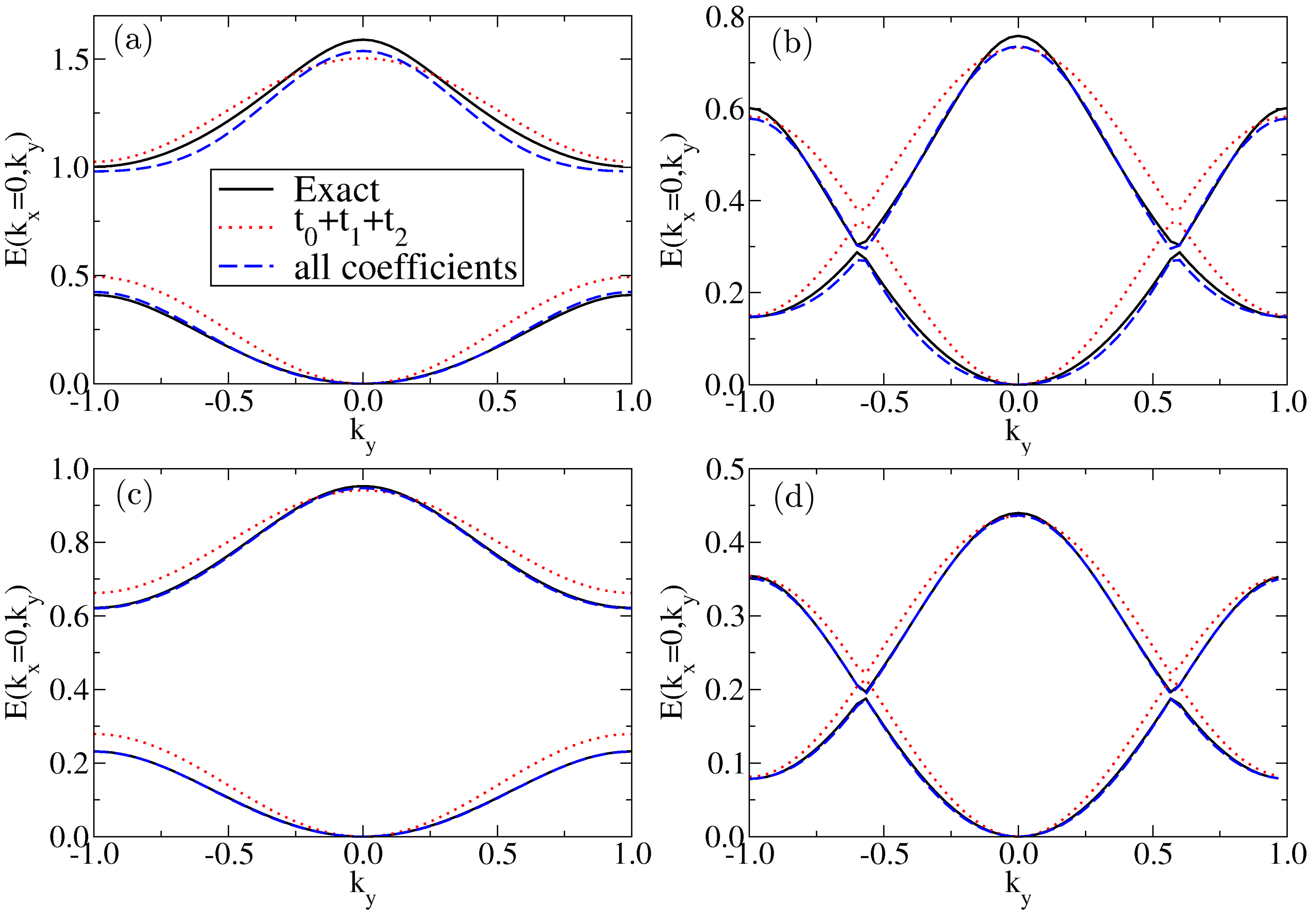}}
\caption{(Color online) Cut of the exact energy bands 
along $k_{y}$ ($k_{x}=0$) compared to the two tight-binding approximations
discussed in the text. 
(a) and (c) represent the dimmer limits
in the experimental and tight-binding regimes, respectively.
(b) and (d) are the analogous in the 1D-chain limit.}
\label{fig:append-bands-approx}
\end{figure}

In Figs. \ref{fig:append-bands-approx}(a),(b) we compare the exact and tight-binding energy
dispersions along $k_{y}$ ($k_{x}=0$) in the dimer and 1D-chain limits (panels (a) and (b), respectively), for the experimental regime \cite{tarruell2012}. 
Here we have included the results for the two tight-binding approximations considered in the text (with just $t_{0}$, $t_{1}$, $t_{2}$, and with all the coefficients).
As in the stretched-honeycomb case (Fig. \ref{fig:sym-bands-apprx}), 
the tight-binding model reproduces the main features of the exact dispersion, including the approximate position of the Dirac point in the case of
Fig. \ref{fig:append-bands-approx}(b) (note that there is no such point in the dimmer limit, Fig. \ref{fig:append-bands-approx}(a)). 
In Figs. \ref{fig:append-bands-approx}(c),(d) we show the analogous pictures for 
the tight-binding regime discussed in the text, that is doubling the potential parameters of Ref. \cite{tarruell2012}. In this case, the agreement with the exact energies when all the coefficients are included is remarkable in both limits.

A further way to test the accuracy of the different tight-binding
expansions is to analyze the overall mismatch 
of the tight-binding energies against the exact ones.
Here, we evaluate this mismatch using the following expression \cite{modugno2012,ibanez2013}
\begin{equation}
\delta E_{1,2}\equiv\frac{1}{\bar{\Delta}}\sqrt{\frac{1}{S_{B}}\int_{\cal{B}}d\bm{k}\left[\varepsilon_{1,2}(k)
-\epsilon_{-,+}(k)\right]^{2}}
\label{eq:mismatch}
\end{equation}
where $\varepsilon_{n}$ are the exact energies, 
${\bar{\Delta}}\equiv (\Delta\varepsilon_{1}+\Delta\varepsilon_{2})/2$
the average bandwidth and $S_{B}$ the area of the Brillouin zone.

\begin{figure}[t!]
\centerline{\includegraphics[width=0.95\columnwidth]{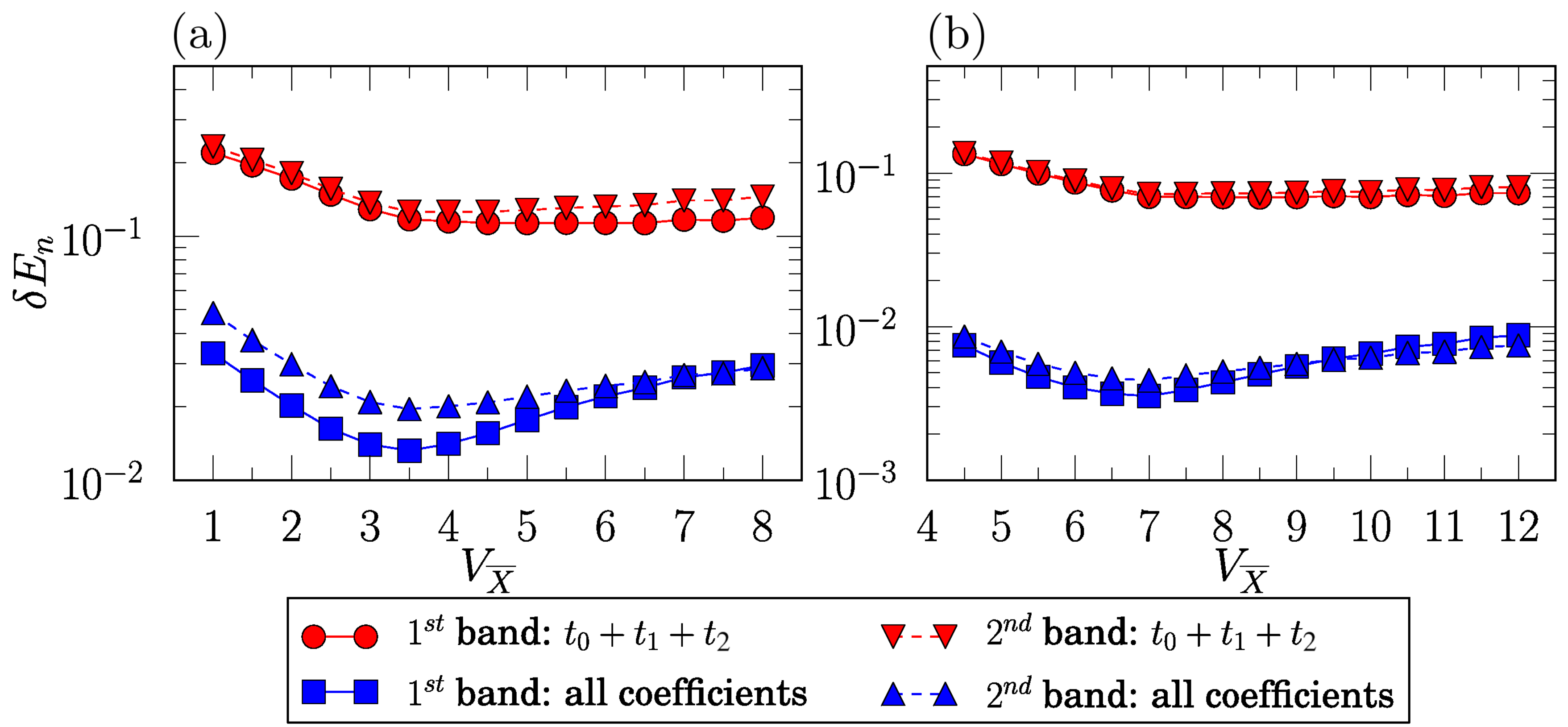}}
\caption{(Color online) Calculated energy mismatch $\delta E_{n}$ for the two bands
including the two tight-binding approximations discussed in the text. 
(a) and (b) respectively show the results for the experimental and tight-binding regimes.}
\label{fig:mistmatch-tarr}
\end{figure}

The calculated mismatch $\delta E_{n}$ is shown in Figs. \ref{fig:mistmatch-tarr}(a),(b) as a function of $V_{\overline{X}}$,  
for the experimental and tight-binding regimes, respectively.
Overall, the mismatch in the tight-binding regime (b) is one order of magnitude smaller 
than the one of the experimental regime (a).
Remarkably, the best approximation in  Fig. \ref{fig:mistmatch-tarr}(b)
has an error below $1\%$ in all the range of $V_{\overline{X}}$.
This further confirms the adequacy of the tight-binding models 
in terms of the MLWFs. 
We identify two  different trends 
in the behavior of the mismatch.
Focusing on Fig. \ref{fig:mistmatch-tarr}(a), we find 
that for $V_{\overline{X}}\lesssim4.5E_{R}$, 
$\delta E_{n}$ decreases as $V_{\overline{X}}$ is increased.
This can be expected, since in this region
the MLWFs become much more localized 
as the potential is raised (see Fig. \ref{fig:spread}), 
hence a more tight-binding regime is approached.
For $V_{\overline{X}}\gtrsim4.5E_{R}$, in contrast, 
the mismatch increases with increasing $V_{\overline{X}}$.
We recall from Fig. \ref{fig:sym-tunnel-tarruel}
that the tunneling coefficients corresponding to sites
inside the 1D-chains grow as $V_{\overline{X}}$ is increased.
When approaching the 1D-chain limit, some of these coefficients 
that are not considered in our tight-binding model
may become relevant, hence the quality of the
approximation may decrease.

\section{Effect of parity breaking}
\label{app:d}
\begin{figure}[b!]
\centerline{\includegraphics[width=0.9\columnwidth]{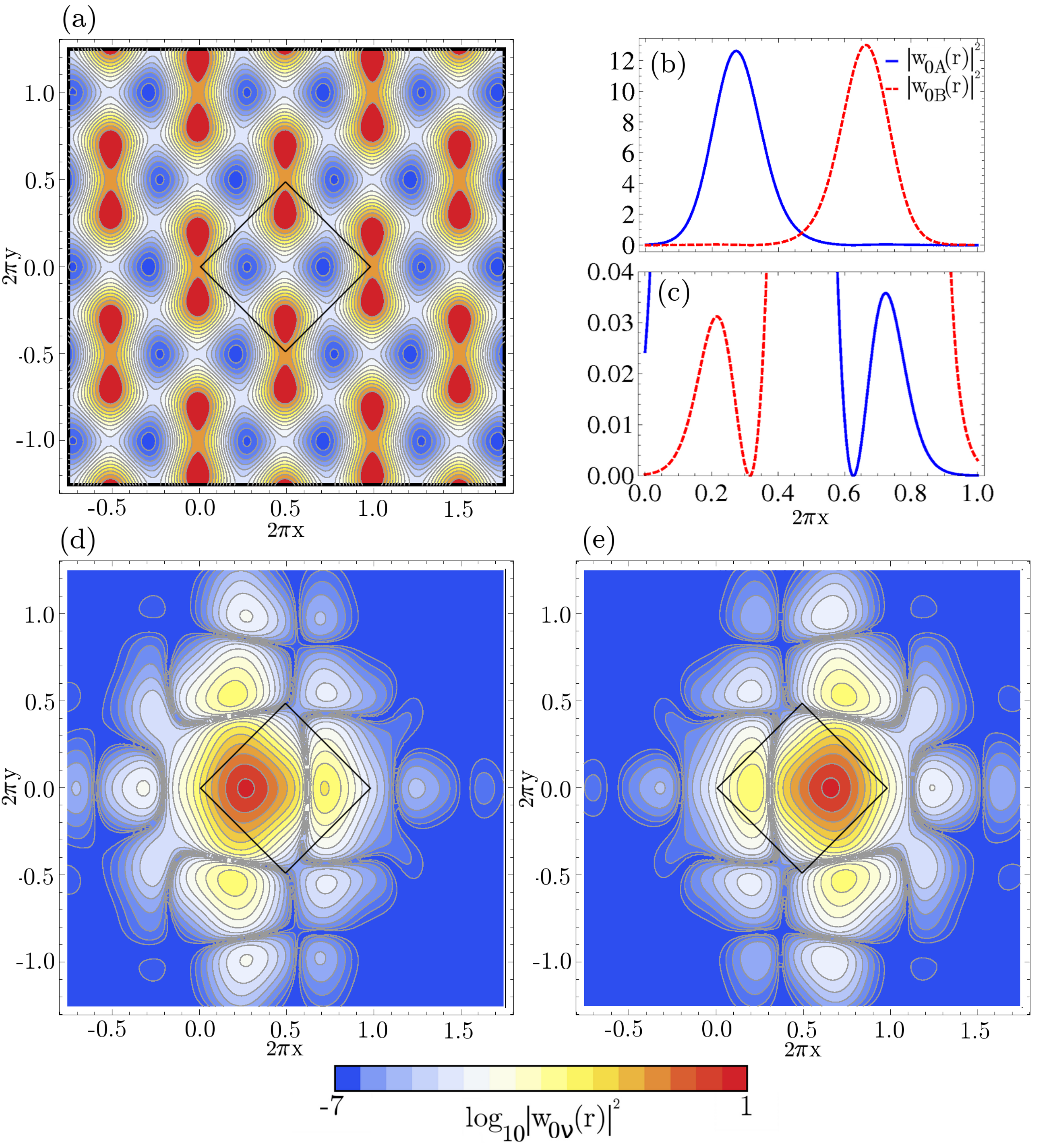}}
\caption{(Color online) Asymmetric structure corresponding to the angle $\theta=\pi+0.1$,
with potential parameters $V_{X}=0.56 $, $V_{Y}=3.6 $, $V_{\overline{X}}=6.94$ (merging point).
(a) illustrates the structure of the potential for this configuration,
showing a deeper minimum at sublattice B than in A 
(color code as in Fig. \ref{fig:sym-potential-mlwf}).
(b) and (c) show one dimensional profiles of $|w_{0A}(x,y=0)|^{2}$
(solid, blue) and $|w_{0B}(x,y=0)|^{2}$ (dashed, red) in the central unit cell. 
Note the different distributions of the
two MLWFs, as a consequence of parity breaking. This is evident also from
the two dimensional plots of $|w_{0A}(\textbf{r})|^{2}$
and $|w_{0B}(\textbf{r})|^{2}$, in (d) and (e) respectively.
}
\label{fig:potential-mlwf-assym}
\end{figure}
\begin{figure}
\centerline{\includegraphics[width=0.9\columnwidth]{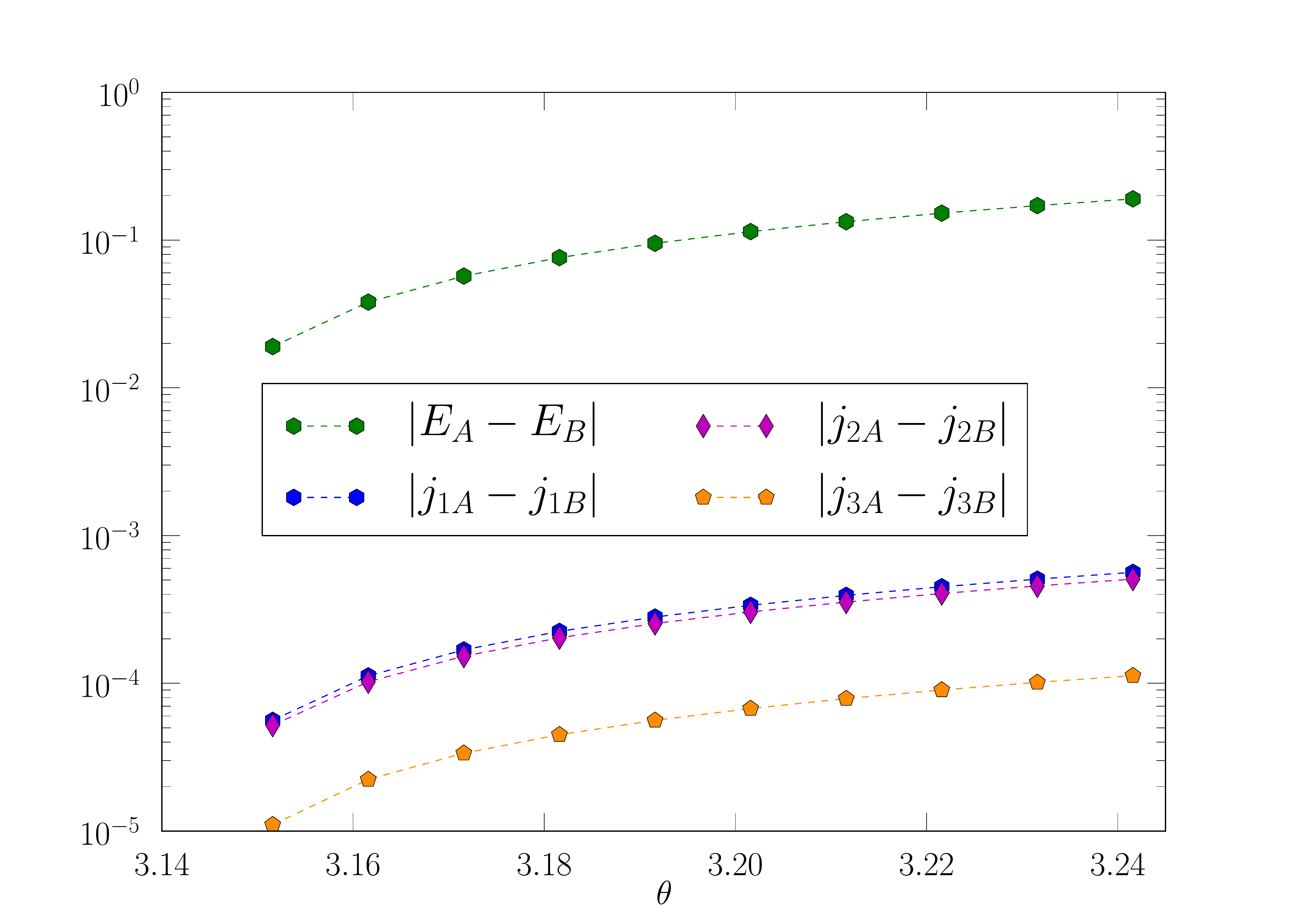}}
\caption{(Color online) Splitting of the diagonal coefficients as a function of the angle $\theta$
(at the merging point: $V_{X}=0.56 $, $V_{Y}=3.6 $, $V_{\overline{X}}=6.94$, cf. Fig. \ref{fig:sym-tunnel-tarruel-str-hon}). Note that $|E_{A}-E_{B}|=2|\epsilon|$, see Eq. (\ref{eq:epsilon}).
}
\label{fig:ym}
\end{figure}

Here we analyze the asymmetric case corresponding 
to $\theta\neq\pi$, considering for simplicity just the tight-binding parameter regime.
In this configuration, the two potential minima in the unit cell
become non-degenerate \cite{tarruell2012}.
In Fig. \ref{fig:potential-mlwf-assym}(a) we illustrate the
structure of the potential for $\theta=\pi+0.1$ at the merging point,
with the deeper minimum at site B.
Correspondingly, the associated MLWFs exhibit a higher localization
around  B, as illustrated in Figs. (b-e).
As a consequence of parity breaking, the degeneracy of the diagonal coefficients is also broken (for both the onsite energies $E_{\nu}=J_{00}^{\nu}$ - see Eqs. (\ref{eq:epsa})-(\ref{eq:epsilon}) - and the tunneling coefficients $j^{\nu}_{i}$, $\nu=A,B$), see Fig. \ref{fig:ym}.
This figure shows the splitting of onsite energies
and diagonal tunneling coefficients for small deviations from $\theta=\pi$
(the off-diagonal tunneling coefficients are weakly affected in these
range of values of $\theta$).
These variations allows to accurately reproduce the
exact dispersion law and in particular the opening of
a mass gap at the Dirac points, as discussed in the text (see Figs. \ref{fig:asym-pi},\ref{fig:asym-075}).

\end{document}